\documentclass[12pt,preprint]{aastex}
\begin{document}

\parindent=1.0cm

\title{THE RECENT STELLAR ARCHEOLOGY OF M31 -- THE NEAREST RED DISK GALAXY \altaffilmark{1}}

\author{T. J. Davidge, A. W. McConnachie}

\affil{Herzberg Institute of Astrophysics,
\\National Research Council of Canada, 5071 West Saanich Road,
\\Victoria, BC Canada V9E 2E7}

\author{M. A. Fardal}

\affil{Department of Astronomy, University of Massachusetts, LGRT 619-E, \\ Amherst, MA 01003-9305, USA}

\author{J. Fliri, D. Valls-Gabaud}

\affil{LERMA, UMR CNRS 8112, Observatorire de Paris, \\ 61 Avenue de l'Obseratoire, 75014 Paris, France}

\author{S. C. Chapman}

\affil{Institute of Astronomy, University of Cambridge, \\ Madingley Road, Cambridge, UK CB3 0HA}

\author{G. F. Lewis}

\affil{Sydney Institute for Astronomy, School of Physics, A28, \\ The University of Sydney, NSW 2006, Australia}

\author{R. M. Rich}

\affil{Division of Astronomy and Astrophysics, University of California, Los Angeles, \\ 430 Portola Plaza, Box 951547, Los Angeles, CA 90095-1547, USA}

\altaffiltext{1}{Based on observations obtained with MegaPrime/MegaCam, a joint
project of CFHT and CEA/DAPNIA, at the Canada-France-Hawaii Telescope (CFHT)
which is operated by the National Research Council (NRC) of Canada, the
Institut National des Sciences de l'Univers of the Centre National de la
Recherche Scientifique (CNRS) of France, and the University of Hawaii.}

\begin{abstract}

	We examine the star-forming history (SFH) of the 
M31 disk during the past few hundred Myr. The luminosity 
functions (LFs) of main sequence stars at distances R$_{GC} > 21$ kpc 
(i.e. $> 4$ disk scale lengths) are matched by models that assume a constant star 
formation rate (SFR). However, at smaller R$_{GC}$ the LFs suggest that 
during the past $\sim 10$ Myr the SFR was $2 - 3 \times$ higher than during 
the preceding $\sim 100$ Myr. The rings of cool gas that harbor 
a significant fraction of the current star-forming activity 
are traced by stars with ages $\sim 100$ Myr, indicating that 
(1) these structures have ages of at least 100 Myr, and (2) stars in these 
structures do not follow the same relation between age and random velocity as 
their counterparts throughout the disks of other spiral galaxies, probably 
due to the inherently narrow orbital angular momentum distribution of the giant 
molecular clouds in these structures. The distribution of evolved red stars 
is not azimuthally symmetric, in the sense that their projected density 
along the north east segment of the major axis is 
roughly twice that on the opposite side of the galaxy. The north east 
arm of the major axis thus appears to be a fossil 
star-forming area that dates to intermediate epochs. Such a structure may be 
the consequence of interactions with a companion galaxy.

\end{abstract}

\keywords{galaxies: individual (M31) --- galaxies: evolution --- galaxies: spiral}

\section{INTRODUCTION}

	As the nearest large spiral galaxy, M31 is a fundamental 
benchmark for studies of disk galaxy evolution, and 
our understanding of the evolution of M31 
has changed profoundly during the past decade. There is a large body of 
evidence (e.g. Ibata et al. 2001, 2007; Hammer et al. 2007; Tanaka et al. 
2010) that interactions with companion galaxies, some of which may not have 
survived intact to the present day and remain only as debris trails, have occured 
throughout the history of M31. The detection of similar tidal features close to 
other nearby galaxies (e.g. Martinez-Delgado et al. 2010) indicates that 
galaxy-galaxy encounters in the local Universe have not been rare, and have played 
a key role in sculpting the current appearance of many nearby galaxies.

	The current study focuses on investigating the star-forming history 
(SFH) of M31 during the past few hundred Myr, 
using the brightest resolved stars as tracers. It has been suggested that 
M31 has interacted with some of its companions during this time, and a key 
element of our work is to search for observational signatures 
of this activity. A review that considers the 
entire history of M31 will guide our understanding of its present-day appearance, 
and allow a conceptual picture of the key events that shaped its evolution 
to be developed. To this end, in the remainder of this section 
we review previous work on the evolution of M31. While current theories for the
buildup of disk galaxies give primacy to the smooth accretion of hot and cold 
gas, the following focuses on the influence of major and minor mergers, as these 
imprint observable signatures on the kinematics, spatial distribution, and 
age distribution of stars. Not every interaction can be traced in 
detail, especially those that happened more than a few dynamical 
times in the past, and the discussion of the earliest phases of the evolution 
of M31 are -- by necessity -- more speculative than the discussion of the 
most recent events. The ages listed in what follows should also 
be viewed as approximations.

\subsection{Initial Assembly (t = 10 Gyr ago)} 

	The initial assembly of M31 likely involved the merger of proto-galactic 
structures, which contained a mix of gas and stars that had formed {\it in situ} 
(e.g. Oser et al. 2010). The gas in these structures probably contributed to the 
assembly of an M31 proto-disk, supplementing material accreted through the 
large-scale inflow of hot and cold gas. Any early disk was likely short-lived 
due to the high frequency of major mergers. In addition to disrupting the 
early disk, mergers may also have spurred the formation of a classical 
metal-poor halo, through the displacement of stars out of the disk plane. 
Stars ejected in such events are expected to dominate the central 20 kpc of halos 
(Zolotov et al. 2009). At larger radii the majority of halo stars 
may not have had such a violent origin, and may instead have been 
accreted from companion galaxies during the first few Gyr of galaxy 
assembly (Font et al. 2008; Zolotov et al. 2009; Cooper et al. 2010). 
A classical metal-poor halo has been detected around M31 (e.g. Kalirai et al. 
2006; Chapman et al 2006; Ibata et al. 2007; Koch et al. 2008). 

	There is evidence that M31 accreted a number of satellites early-on. 
The integrated luminosities, masses ($\sim 10^9$ M$_{\odot}$), and dynamical 
properties of the M31 and the Galactic halos are similar (Chapman et al. 2006; 
Ibata et al. 2007), and the properties of Galactic globular clusters suggest 
that 6 -- 8 satellites large enough to form such large star clusters 
may have been accreted (e.g. MacKey \& Gilmore 2004; Forbes \& Bridges 2010). 
Globular clusters in the M31 halo tend to be associated with tidal debris 
(Mackey et al. 2010), and the range of ages, chemical mixtures, and overall 
metallicities amongst M31 clusters (e.g. Beasley et al. 2005) 
hints at diverse progenitors. Some objects that were originally 
identified as globular clusters may not be simple 
stellar populations (SSPs) (e.g. Davidge et al. 1990; Meylan et al. 2001, 
Fuentes-Carrera et al. 2008), but instead may be the remnants of dwarf galaxies 
that were shredded early in the evolution of M31 (Fuentes-Carrera et al. 2008). 

	The properties of classical globular clusters suggest that 
M31 and the Galaxy may have experienced different 
chemical enrichment histories early-on. When compared with Galactic 
globular clusters, the spectra of many M31 globular clusters have relatively strong 
CN absorption bands (e.g. Burstein et al. 1984; Davidge 1990a), and this has been 
attributed to a comparatively large nitrogen abundance in M31 clusters 
(Burstein et al. 2004). Such an abundance difference may have its origins in 
the numbers of present-day stars in globular clusters that formed in 
the diffuse proto-cluster environments, as opposed to those that formed in 
the more compact potential wells that existed 
after cluster formation (Carretta et al. 2010), in the sense that M31 
clusters are made up of a smaller fraction of progenitor stars. If correct, then 
the location of stars in M31 globular clusters on the [Na/Fe] versus [O/Fe] 
diagram, which is a diagnostic of the rate of chemical enrichment, should differ 
from that defined by Galactic objects.

	Early merger activity likely also 
produced a pressure-supported bulge (e.g. Bekki \& Chiba 2001). 
Aside from the central few arcsec, the integrated spectrum of 
the M31 bulge at visible wavelengths originates predominantly from old 
stars (e.g. Davidge 1997; Puzia et al. 2005; Saglia et al. 2010). 
The brightest AGB stars are well-mixed throughout the 
central $\sim 1$ kpc of the M31 bulge, as expected if they formed during a 
rapid, uniform star-forming episode (Davidge 2001b), such as would result from 
the violent merging of satellites. Still, the structural (e.g. Beaton 
et al. 2007) and kinematic (e.g. Morrison et al. 2010) properties of the 
M31 bulge suggest that secular processes may have also contributed to its growth.

	The material from which stars in the Galactic bulge formed may have been 
chemically enriched by the earliest generation of globular clusters (Davidge 
2001a). If globular clusters did play a significant role in enriching the early ISM 
of M31 then the chemical properties of stars in the bulge and globular clusters may 
show similarities. In fact, the strengths of CN absorption bands at visible 
wavelengths in the integrated spectrum of the inner bulge of M31 are reminiscent 
of those in M31 globular clusters (Davidge 1997), suggesting a chemical kinship.

	A thick disk may also have been formed during the early assembly of M31. 
Using kinematic selection criteria, Collins et al. (2011) detect a thick disk in 
M31 and find that the component stars are $0.2 - 0.3$ dex 
more metal-poor than thin disk stars at the same galactic radius. The estimated 
total mass of the thick disk suggests that it did not form from the 
destruction of a satellite that follows the trend between [M/H] and total 
mass defined by present-day systems. If the M31 thick disk formed from 
thin disk stars that were kinematically heated then its (comparatively) low 
metallicity suggests that this happened during early epochs, before the major 
merger that produced the extended extraplanar component discussed in the 
next section.

	The spatial extent of M31 at the end of its initial assembly 
was significantly smaller than the present-day galaxy. Simulations 
suggest that a galaxy with a present-day mass like that of 
M31 probably accreted no more than one-half of this mass by moderate redshifts 
(Stewart et al. 2008), while at $z \sim 2$ the spatial extent of its disk 
may have been substantially (by a factor of $\sim$ two-thirds) smaller than its 
present-day value (Firmani \& Avila-Reese 2009). The modest size of the 
M31 progenitor aside, it was this system that served as the seed for the subsequent 
evolution that lead to the present-day galaxy. 

\subsection{The Formation of the Present-Day Disk (t = 8 Gyr ago)} 

	Mergers were common events during intermediate epochs, and 
played a key role in sculpting the morphological properties of nearby 
disk galaxies (Hammer et al. 2007; 2009; van der Kruit \& Freeman 2011). 
Simulations of a merger between a disk galaxy and a large companion find that the 
existing disk can be obliterated, with the orbits of disk stars thermalising 
to form a pressure-supported extraplanar population (e.g. Hopkins et al. 2009). 
A gas disk will re-form if the progenitors have sufficient 
quanities of gas (Robertson et al. 2006; Governato et al. 2007).

	There is evidence that a violent merger occured only a few Gyr following 
the initial assembly of M31. Such an event would have profoundly affected the 
morphology of M31, and may have produced features that are usually attributed 
to more recent interactions (Hammer et al. 2010). The proposed merger heated the 
disk of M31 and produced a diffuse extraplanar component that is distinct from 
the classical metal-poor halo. It has long been known that the extraplanar 
regions of M31 are dominated by stars with metallicities that (1) are much higher 
than those in the Galactic halo (e.g. Mould \& Kristian 1986, 
Pritchet \& van den Bergh 1988, Durrell et al. 2001) and (2) are consistent 
with an origin in either a large disk or an SMC/LMC-like companion. More recently, 
a kinematically hot component (e.g. Chapman et al. 2006; Gilbert et al. 2007) has 
been detected, that has been traced out to extraplanar distances of 60 kpc. 
This component is distinct from the thick disk, which has a vertical scale height 
of 3 kpc (Collins et al. 2011). The metallicities of stars in this component show 
little or no trend with distance from the galaxy 
(Chapman et al. 2006; Koch et al. 2008; Richardson et al. 2009). 
A break in the surface brightness profile occurs near 30 kpc 
(Ibata et al. 2007), possibly indicating the transition between a violently 
disturbed inner halo and a more metal-poor classical 
halo. The lack of sub-structure in the distribution of the 
moderately metal-rich extraplanar stars suggests that they are not the remnants of 
a galaxy that was slowly accreted by M31, but instead are the result of a 
kinematically violent event.

	The extraplanar component must have an age of at least 
a few Gyr, as the C/M5+ ratio is $0.10 \pm 0.05$ throughout the 
outer regions of M31 (Koch \& Rich 2010). Such a C star frequency amongst stars 
with a moderately sub-solar metallicity is suggestive of an 
age in excess of $\sim 3 - 6$ Gyr (e.g. Cole \& Weinberg 2002).
Brown et al. (2008) discuss observations of extraplanar fields with 
minor axis distances of 21 and 35 kpc, which are populated by stars with 
main sequence turn-off (MSTO) ages $\geq 8$ Gyr.\footnote[2]{A field that 
is 11 kpc above the disk was also observed by Brown et al. (2008). The stars 
in that field tend to have ages $\geq 4$ Gyr, with a modest number of 
younger objects. The majority of stars in the 11 kpc field are probably not part 
of a pressure-supported system, but are likely associated with the disk (e.g. Ibata 
et al. 2007).} The age distribution is such that the majority of stars in 
the 35 kpc field are not associated with the early assembly of M31, as only 10\% 
are extremely old and metal-poor and so likely belong to the classical halo. 
That the bulk of the stars in the two Brown et al. (2008) fields have 
ages $> 8$ Gyr argues that (1) the merger occured a few Gyr after the 
initial assembly of M31, and (2) there has not been a subsequent merger of 
comparable magnitude -- while M31 has likely been subject to encounters with other 
galaxies in the more recent past (see below), none of these have been as disruptive 
as the event described here. 

	A SFR for M31 during this epoch can be estimated
from the relation between the brightness of the brightest globular cluster, 
normalized to an age of 10 Myr, and SFR. The most luminous globular cluster that 
is known to have formed during intermediate epochs is 292-010, for which Beasley et 
al. (2004) find a mass of $1.7 \times 10^5$ M$_{\odot}$. The models 
of Bruzual \& Charlot (2003) indicate that log(M/L)$_V \sim -1.5$ for a population 
with an age 10 Myr, and so 292-010 would have had M$_V = -16.8$ when it 
was 10 Myr old. Applying relation 1 from Bastian (2008) yields a SFR of 
$\sim 320$ M$_{\odot}$ year$^{-1}$.

	Major mergers trigger large-scale star-forming 
activity that is not restricted to the central regions of the galaxy (e.g. 
Teyssier et al. 2010), and M31 may have appeared as an ultra-luminous 
infrared galaxy (ULIRG) for a period of time during intermediate epochs. Star 
formation is seen throughout the disks of ULIRGs, and is curtailed at progressively 
smaller radii as torques cause gas to move inwards (e.g. Soto \& Martin 
2010). If M31 was a typical ULIRG then 5 -- 10\% of its current stellar mass may 
have been produced over a $\sim 0.1$ Gyr (Marcillac et al. 2006) time span, 
and this is roughly consistent with the SFR estimated above from the brightest 
intermediate age globular cluster.

	The formation of globular clusters in the local Universe is associated 
with the large-scale re-distribution of gas that occurs during major
mergers (e.g. Whitmore et al. 1993), and the concomitant violent turbulent stirring 
of the ISM that produces dense star-forming clumps (Teyssier et al. 2010). 
Globular clusters in M31 span a wide range of ages (e.g. Beasley et al. 2004; 
Burstein et al. 2004; Fusi Pecci et al. 2005; Caldwell et al. 2009), 
and Puzia et al. (2005) find peaks in the cluster age distribution at $t \sim 
12$ Gyr, 8 Gyr, and 1 Gyr. The majority of globular clusters 
in M31 belong to the oldest age group, which is associated with the initial 
assembly of the galaxy. However, the number of clusters that belong to the 
8 Gyr peak -- and which likely formed during the major merger that produced 
the extraplanar component -- is a significant fraction of the older 
cluster system. These intermediate age clusters tend to be seen 
against the disk of the galaxy, and are not at 
large radii. To the extent that globular clusters are probes of a disturbed and 
turbulent ISM, the 8 Gyr cluster population is the remnant of an event that 
was almost as violent as the mergers associated with the initial assembly of M31.

	A major merger would have imprinted signatures on the age distribution of 
stars throughout much of the disk. If the present-day disk of M31 formed 
during intermediate epochs following the obliteration of an 
earlier disk then there should be (1) an absence of disk stars with 
ages $> 8$ Gyr, and (2) a large population of disk stars that formed 
$\sim 8$ Gyr in the past - these are the field counterpart of the globular cluster 
population that formed at that time. In addition, starbursts deplete the gas supply 
at a rate that is an order of magnitude shorter than in a normal disk (Daddi 
et al. 2010). Therefore, the SFR would have plunged within a few disk rotation 
times of the merger. 

	Observations that fully test these predictions require photometric 
depths and angular resolutions that tax the capabilities of existing telescopes. 
Still, Olsen et al. (2006) investigate the SFH of bulge and disk fields in M31, and 
find a preponderance of stars with ages $\geq 8$ Gyr. This is consistent 
with the disk of M31 experiencing a large burst of star formation in the past 
that may not have been associated with the epoch of earliest disk assembly.

\subsection{A Close Encounter with M33 (t = 2 - 4 Gyr ago)}

	M31 has not evolved in isolation since the major merger that 
spurred the formation of the present-day thin disk, and there 
is a high probability that there was a close encounter between 
M31 and M33 within the past few Gyr. Various studies have estimated when 
such an interaction may have occured. The SFH of M33 suggests that the encounter 
must have happened at least 0.5 Gyr in the past (Davidge \& Puzia 2011). Using the 
locations and velocities of M31 and M33, Putman et al. (2009) argues that an 
encounter occured 1 -- 3 Gyr in the past, while Bekki (2008) assigns an 
age of 4 -- 8 Gyr based on the properties of the HI bridge that may 
link the galaxies (Braun \& Thilker 2004). Simulations discussed by McConnachie et 
al. (2009) that use currently available orbital constraints and the 
assumption that M33 was significantly perturbed by the encounter 
place this interaction a few billion years in the past.

	There is a high probability that the tidal radius of M33 was $\leq 15$ 
kpc during this encounter, and this would have disturbed the disk of M33. 
The S-shaped HI distribution in the outer regions of M33 is testiment of such a 
disturbance, and this structure likely contains gas that was stripped from 
the M33 disk (Putman et al. 2009). The detection of extended stellar structures 
in the outer regions of the M33 disk, coupled with isophotal distortions at 
smaller radii, provide further evidence that the disk of M33 was disturbed 
(McConnachie et al. 2010; Davidge \& Puzia 2011).

	The disk of M31 would also have been affected by this event 
(McConnachie et al. 2009; Davidge \& Puzia 2011), with elevated 
levels of star-forming activity being sparked as tidal forces re-distributed gas. 
Intermediate age globular clusters in M31 with ages $\sim 2$ Gyr (Puzia et al. 
2005) may have formed during this event. While the SFR of the M31 disk would 
have been greatly elevated immediately after the encounter, it would 
eventually drop as the ISM is depleted and disrupted. 
There is evidence of a decrease in M31 star-forming activity 
$\sim 1$ Gyr ago (Williams 2002). Given that the damping 
timescale for elevated SFRs is on the order of a few disk crossing times, then 
the drop in the SFR found by Williams (2002) is consistent with an interaction that 
had peak star-forming activity $\sim 2$ Gyr ago.

	If gas was re-distributed throughout M31 then intense 
star-forming activity would likely have been triggered in the galaxy center, 
and this part of M31 {\it does} contain stars that formed at 
a time that coincides with the proposed interaction with M33. The integrated 
spectrum of the central few arcsec of M31 at visible 
wavelengths reveals a large population of objects with an age of a few Gyr 
(Davidge 1997; Sil'chenko et al. 1998). Saglia et al. (2010) conclude 
from integrated spectra that the luminosity-weighted age of stars near 
the center of M31 is 4 -- 8 Gyr.

\subsection{The Recent Past (t $< 2$ Gyr ago)}

	Mergers and interactions with companions that have $\leq 10\%$ the mass of 
the primary galaxy are not expected to permanently disrupt disks (e.g. Stewart et 
al. 2008), and there is evidence that M31 has interacted with companions in this 
mass range during recent epochs. The most overt signature of such an encounter may 
be the Giant Stellar Stream (GSS; Ibata et al. 2001). The GSS is a substantial 
structure, with the northern and southern ends separated 
by at least 130 kpc along the line of sight (McConnachie et al. 2003).

	The stellar content of the GSS suggests that the progenitor was a massive 
dwarf galaxy, and Ibata et al. (2001) discuss NGC 205 and M32 as possible sources. 
The photometric properties of RGB stars in the GSS are indicative of a 
moderately metal-rich system, with a 0.5 dex dispersion in [M/H] (McConnachie 
et al. 2003). Using the red clump of core helium-burning stars as an age indicator, 
Tanaka et al. (2010) conclude that stars in the GSS have an age near 7 Gyr, and 
a peak [M/H] near $-0.3$. This metallicity is comparable to that of stars 
in the LMC, and suggests a progenitor mass $\sim 10^9 - 10^{10}$ M$_{\odot}$. 
Fardal et al. (2007) and Ibata et al. (2007) arrive at similar 
progenitor mass estimates. Upper limits to the mass of the progenitor can also be 
placed. Assuming that the GSS progenitor passed through the disk 
of M31, then Mori \& Rich (2008) conclude that it must have had a mass 
$< 10^{10}$ M$_{\odot}$ so as not to widen the disk of M31 more than is observed.

	Font et al. (2006) conclude that the encounter that formed the 
GSS occured only a few hundred Myr in the past, while Mori \& Rich (2008) 
assign an age $\sim 1$ Gyr. The age predicted for the GSS by Font 
et al. (2006) is similar to that modelled by Block et al. (2006) for the 
supposed passage of a satellite (presumably M32) through the M31 disk.
The mean metallicity of stars in the outer regions of M32 is $\sim -0.25$ 
(Rose et al. 2005), which is consistent with the metallicity of stars in the GSS. 

	Models that attempt to reproduce the orbital properties 
of the GSS predict that the encounter occured much more recently than might 
be inferred from its stellar content. The timings predicted by kinematic 
studies and the ages of stars in the GSS are only in conflict 
if the progenitor contained a reservoir of cool gas that 
could have fueled star formation up to the time of its disruption. If the 
progenitor was depleted of gas a few Gyr in the past -- and most present-day 
companions of M31, such as M32, are gas-depleted (e.g. Welch, Sage, \& Mitchell 
1998; Sage, Welch, \& Mitchell 1998; Grcevich \& Putman 2009) -- then 
there is a clear physical explanation for the apparent 
conflict between the kinematic and stellar content age estimates.

	The extraplanar regions of M31 are laced with stellar 
streams and tidal structures, some of which may or may not be related to the GSS. 
The completeness of the catalog of these structures is uncertain, as there 
is a bias against the detection of metal-poor and/or old features. This bias 
occurs because higher surface brightness structures tend to have 
higher metallicities and/or formed more recently (e.g. Gilbert et al. 2009). 

	The diverse stellar contents of streamlets along the minor axis of M31 
led Ibata et al. (2007) to suggest that these structures have multiple 
progenitors, all of which -- based on metal content -- would have initially been 
larger than the Fornax dwarf spheroidal. Mori \& Rich (2008) find that 
a single minor merger can pollute a large fraction of the 
extraplanar regions of M31 within $\sim 2$ Gyr of an encounter. Still, they 
also find that some structures in the outer regions of M31 likely formed more than 
3 -- 4 Gyr in the past, and so are not related in a direct way to the GSS, to which 
they assign an age of 1 Gyr. Fardal et al. (2008), Tanaka et al. (2010), and 
Hammer et al. (2010) argue that many features in the outer regions of M31 
could result from an interaction between M31 and a single galaxy. 
The stream-to-stream diversity in stellar contents indicates that a progenitor 
that produces multiple observed streams would likely have had 
a metallicity gradient, and such gradients are seen in nearby Local Group 
dwarf galaxies (e.g. Harbeck et al. 2001). 

	Clues of a significant accretion event within 
the past few hundred Myr also lurk in the central regions of M31. 
The nuclear star cluster P3 has an age of 200 Myr (Bender 
et al. 2005), and hence formed at the same time as the events 
modelled by Block et al. (2006) and Font et al. (2006). Davidge et al. (2006) find 
a bright object that is a few arcsec from the nuclear star cluster P3. This source 
dominates the central light output of M31 at $4.5\mu$m and 
appears to be a massive dust-enshrouded AGB star that -- based on its luminosity -- 
may have an age of a few hundred Myr. Saglia et al. (2010) find a 
counter-rotating ring of gas in the central regions of M31, which they interpret as 
the remnant of a merger $\sim 0.1$ Gyr in the past. They suggest that $10^6$ 
M$_{\odot}$ of stars formed in the central parts of M31 following this merger.

	The disk of M31 also contains signatures of an event (or events) that 
disrupted the ISM. The HI disk of M31 is twisted and warped (e.g. Corbelli et al. 
2010), and there is a significant non-rotational kinematic component (e.g. Unwin 
1983). Yin et al. (2009) model the chemical properties and the distribution of gas 
and stars in M31 assuming evolution in isolation. 
The model does not match the observed gas content at 
large radii or the radial distribution of star-forming 
activity throughout the galaxy. Yin et al. (2009) attribute these discrepancies 
to interactions. An alternative is to accept that the star-forming efficiency in 
M31 varies with radius, but stays constant throughout the Galaxy 
(e.g. Marcon-Uchida et al. 2010).

	The location and distribution of star-forming regions in M31 also differ 
systematically from those in the Galaxy and M33. While star formation in the 
Galaxy and M33 tends to be distributed along well-defined spiral arms, a large 
fraction of the star-forming activity in M31 occurs in 
rings at R$_{GC} = 10$ and 14 kpc (Thilker et al. 
2005; Gordon et al. 2006; Barmby et al. 2006), in arcs, and in spiral arm 
fragments (e.g. discussion in Efremov 2010). Much of the star-forming activity 
is (and has been for at least the past 100 Myr - \S 6) concentrated in the ring at 
R$_{GC} = 10$ kpc, which contributes substantially to the 
total H$\alpha$ emission from M31 (Devereux et al. 1994). 
Despite the evidence that the ISM of M31 has been disrupted, 
the star-forming environment in these rings is probably not too different 
from that in the Galactic disk. Indeed, GMCs in M31 fall along the same size versus 
line width relation as Galactic GMCs (Sheth et al. 2008), indicating 
a structural kinship.

	The depletion timescale of molecular gas indicates 
that M31 is much less efficient at forming massive 
stars than M33 (Tabatabaei \& Berkhuijsen 2010). The strength 
of the UV radiation field throughout the M31 disk is only 40\% that in the 
solar neighborhood (Montalto et al. 2009), and the SFR of M31 is roughly one half 
that expected for a galaxy of similar mass and morphology that has evolved in 
isolation (Yin et al. 2009). Williams (2003) infers a SFR of 
$\sim 1$ M$_{\odot}$ year$^{-1}$ for M31, based on studies of resolved stars 
over a 1.4 degree$^2$ area. Integrated light diagnostics suggest present-day 
SFRs for M31 of 0.3 (Tabatabaei \& Berkhuijsen 2010), 0.4 (Barmby et al. 2006) 
and 0.6 (Kang et al. 2009) M$_{\odot}$ year$^{-1}$. These are lower 
than the SFR of the Galaxy (e.g. Boissier \& Prantzos 1999; Robitaille \& Whitney 
2010) and are on par with the much less massive galaxy M33 (Hippelein et al. 2003; 
Verley et al. 2007; Verley et al. 2009).

\subsection{The Present Study}

	In the present paper, deep MegaCam observations are used to 
study the locations and photometric properties of bright resolved stars 
throughout the disk of M31. Massive main sequence and evolved stars 
are detected, and these are used to characterize the SFH during 
the past few hundred Myr and search for signatures of recent interactions. 
Atmospheric seeing and sky emission restrict the 
photometric depth that can be achieved with ground-based datasets when compared 
with space-based observations. However, ground- and space-based datasets of 
nearby galaxies provide complementary information because they usually sample very 
different spatial scales. The spatial coverage of space-based datasets 
of M31 at visible and near-infrared wavelengths tend to be 
limited. The most extensive such dataset is that discussed by Dalcanton et al. 
(2011), which is restricted to the north east quadrant of the galaxy. 
In contrast, our MegaCam data cover almost the entire star-forming 
disk of M31 out to major axis distances in excess of 30 kpc ($> 5$ 
scale lengths). Such comprehensive spatial coverage is of interest 
since the SFH can vary with location over large spatial scales in disks (e.g. 
Davidge 2010; Davidge \& Puzia 2011). Spatial variations of this 
nature can complicate efforts (1) to determine a representative global SFH if 
only part of the disk is sampled, and (2) to assess the large 
scale radial trends that are probes of galaxy evolution. 

	Two important considerations are the distance to M31 and the reddening 
model. An absolute distance modulus $\mu_0 = 24.36$, computed by Vilardell et al. 
(2010) from two double-lined spectroscopic eclipsing binaries, 
is adopted for this study. Such systems are primary 
distance indicators (de Vaucouleurs 1978), that rely on geometric 
parameters directly measured from the application of basic physics. 
Distances measured with these objects should be less susceptible to 
calibration uncertainties than those computed from secondary distance indicators. 
The distance modulus adopted here is in excellent agreement with that computed from 
Cepheids by Riess et al. (2012).

	The baseline reddening model consists of a 
linear combination of foreground (A$_B = 0.27$; Schlegel et al. 1998) 
and internal (A$_B = 0.61$; Pierce \& Tully 1992) components. 
The total extinction is thus A$_B = 0.88$ magnitudes. This corresponds to 
A$_{u'} = 1.05$, A$_{g'} = 0.77$, and A$_{r'} = 0.56$ magnitudes, based on 
the relations listed in Table 6 of Schlegel et al. (1998). 
It is demonstrated in \S 4 that this total extinction holds for a `typical' 
main sequence star in our sample. Of course, the reddening towards any given 
object depends on factors such as its age, location, and 
evolutionary state (e.g. Massey et al. 2009), and it is not 
surprising that there is a substantial spread in extinction 
throughout the M31 disk. For example, Hodge et al. (2010) find a dispersion of 
$\pm 0.23$ mag in $E(B-V)$ measurements among young clusters. 
There is almost certainly an environmental component, as 
Hodge et al. (2010) and Perina et al. (2010) find $E(B-V) = 0.28$ (A$_{g'} = 
1.06$) for young star clusters, whereas Massey et al. (2007) measure 
$E(B-V) = 0.13$ (A$_{g'} = 0.49$) for young field stars in the M31 disk. 

	The paper is structured as follows. The observations and 
the reduction of the data are discussed in \S 2, while details of the 
photometric measurements, including their calibration and characterization, are 
discussed in \S 3. Color-magnitude diagrams (CMDs) 
are presented in \S 4, while the SFH of the M31 disk 
is investigated in \S 5 using the LFs of main sequence stars. The spatial 
distribution of main sequence and evolved stars is examined in \S 6. 
The paper closes with a summary and discussion of the results in \S 7.

\section{OBSERVATIONS \& REDUCTIONS}

	The data were recorded on the 3.6 metre CFHT 
as part of the 2010B MegaCam (Boulade et al. 2003) observing queue. The 
detector in MegaCam is a mosaic of thirty six $2048 \times 4612$ E2V CCDs. These 
are deployed in a $4 \times 9$ format, and cover roughly one 
degree$^2$ exposure$^{-1}$ with 0.185 arcsec pixel$^{-1}$.

	Four fields that cover almost all of the star-forming disk of M31 
were imaged through $u^*, g'$, and $r'$ filters.\footnote[3]
{The filter transmission curves can be found at 
\newline http://www.cfht.hawaii.edu/Instruments/Filters/megaprime.html} 
The locations of the MegaCam pointings are 
shown in Figure 1. The observations for each field were obtained during a 
single continuous block of time, that typically spanned 2 hours. 
While archival $g'$ and $r'$ MegaCam data were available of the disk of M31, 
images in these filters were still recorded for our program 
to secure a set of single epoch measurements, and thereby 
suppress scatter in colors that might arise from photometric variability 
over timescales $\geq 2$ hours. A log of the observations, showing dates of 
observations and exposure times for each field, is shown in Table 1. 
The $u^*$ images have markedly longer exposure times 
than the $g'$ and $r'$ images because of the comparatively low 
system throughput at shorter wavelengths.

	Initial processing of the data was done 
with the CFHT ELIXIR pipeline, and this included 
bias subtraction and flat-fielding. Each pipeline-processed image 
was divided into six $2 \times 3$ CCD sub-mosaics to produce datasets of a 
manageable size for subsequent processing and photometric analysis. 
Sub-mosaic images were spatially registered to correct for offsets 
introduced during acquisition, and the results 
were stacked according to field and filter.

	The MegaCam optics produce modest chromatic 
distortions across the imaged field, which manifest as location-dependent offsets 
between the centroids of stars imaged in different filters. The offsets 
between $u^*$ and $g'$ images taken at moderate airmass and 
aligned near the optical axis of the instrument can be up to an arcsec
near the edge of the MegaCam science field. To correct for these, 
the $u^*$ and $r'$ data were mapped into the $g'$ reference frame 
using the IRAF GEOMAP/GEOTRAN tasks. Stars in the final distortion-corrected 
images typically have FWHM $= 0.7 - 0.9$ arcsec. 

\section{PHOTOMETRIC MEASUREMENTS AND CHARACTERIZATION}

\subsection{Basic Methodology}

	Photometric measurements were made with the point spread function 
(PSF)-fitting program ALLSTAR (Stetson \& Harris 1988). Each PSF was constructed 
from between 50 to 100 bright, unsaturated stars using routines in the DAOPHOT 
package (Stetson 1987). In addition to brightness, PSF stars were selected 
using star-like appearance and the absence of bright neighbors as 
criteria. Faint companions were subtracted from the wings of PSF 
stars in an iterative manner, using progressively improved PSFs.

	Following the procedure described by Davidge (2010), the 
raw photometric catalogues produced by ALLSTAR 
were culled of extended and very faint sources using the error 
in magnitude that is computed by ALLSTAR, $\epsilon$. This quantity 
measures the quality of the PSF fit, but does not take into account other 
sources of error, such as those introduced by crowding. While $\epsilon$ 
is only a lower limit to the total uncertainty in source brightness, it
provides a means of identifying objects for which photometry may be problematic. 

	Two rejection criteria were applied. First, all sources with $\epsilon 
\geq 0.3$ mag were deleted. Obtaining meaningful photometric measurements is 
problematic for objects with such large uncertainties, the majority of which 
are near the faint limit of the data. Second, point sources 
define a sequence in the magnitude {\it vs.} $\epsilon$ plane that is narrow 
at the bright end and fans out towards fainter magnitudes. 
Objects that are outliers from this trend were removed. This step tends to 
remove extended sources (e.g. Davidge 2010) and cosmetic defects.

\subsection{Calibration}

	Standard stars are observed during each MegaCam observing block. These 
observations are used to determine photometric zeropoints, and the results are 
placed in MegaCam image headers during ELIXIR processing. We used these header 
entries to transform instrumental magnitudes into the Sloan Digital 
Sky Survey $u'g'r'$ system (Fukugita et al. 1996). We note that a 
comparison of photometric measurements in overlapping 
sections of fields indicates that the internal field-to-field calibration 
consistency is a few hundredths of a magnitude.

	There are significant differences between the $u^*$ and $u'$ bandpasses, 
with the central wavelength of $u^*$ falling 200\AA\ redward of $u'$. 
Our $u'$ magnitudes were checked using the $UVR$ measurements of bright 
stars in M31 published by Massey et al. (2006). The magnitudes from that study were 
transformed into the Sloan system using equations from Smith et al. (2002), and 
the results were compared for stars in common with our dataset. 

	The comparisons were restricted to bright stars in uncrowded parts of 
M31 to suppress the influence of image quality differences 
between the KPNO and CFHT observations. One consequence of restricting the areal 
coverage in this way is that the majority of the stars used in the comparison have 
red colors, as the main body of the star-forming disk (where bright blue stars tend 
to be found) is intentionally avoided. This caveat notwithstanding, 
there is excellent agreement between the two sets of measurements, with 
$\Delta g' = 0.035$ and $\Delta u' = -0.034$, in the 
sense CFHT -- KPNO. The standard deviations about the mean are 
$\sigma_{\Delta g'} = \pm 0.011$ and $\sigma_{\Delta u'} = \pm 0.079$. 
This agreement is comparable to that found by Davidge \& Puzia (2011) in their 
study of M33.

\subsection{Photometric Characterization: Artificial Star Experiments}

	Sample completeness and uncertainties in the photometry were assessed 
through artificial star experiments. Artificial stars were assigned magnitudes and 
colors that hold for main sequence stars, which are 
the primary probes of the SFH in this paper. 
As with observations of real objects on the sky, an 
artificial star was considered to be recovered only if (1) it was detected in two 
filters (either $u' + g'$ or $g' + r'$), and (2) it survived 
the $\epsilon$--based culling criteria described in \S 3.1.

	The uncertainties and systematic effects in the 
photometry that arise due to crowding and/or statistical fluctuations in the noise 
increase substantially with magnitude when sample completeness 
drops below 50\%. The magnitude at which 50\% 
completeness occurs is thus one estimate of the faint limit. 
The artificial star experiments indicate that 50\% 
completeness occurs near $u',g' = 25.5$ and $r' = 24.5$ in the outer regions 
of the disk, and $u',g' = 25$ and $r' = 23.8$ at intermediate 
radii along the major axis. These values are roughly consistent with the 
observed faint limits of the CMDs (\S 4).

\subsection{Photometric Characterization: Image Stacking Experiments}

	The impact of crowding on photometry can be investigated 
by stacking images to simulate higher stellar densities (e.g. Davidge 2001b). A 
shortcoming is that foreground stars and background galaxies, 
which are more-or-less uniformly distributed on the sky over angular scales of a 
few degrees, are over-represented in the stacked images. This can be 
mitigated by selecting fields for stacking that have a moderately high initial 
stellar density, with the aim of minimizing the number of frames that are 
co-added to simulate a given stellar density.

	Four $500 \times 500$ pixel$^2$ areas from Field 1 were selected as the 
building blocks for the stacking analysis. These are $\sim 38$ arcmin from the 
center of the galaxy, where the average surface brightness is $\mu_B = 22.8$ 
magnitudes arcsec$^{-2}$ (Walterbos \& Kennicutt 1987). 
The faint limit of the CMDs in this part of the disk 
is almost the same as the outermost parts of the disk (\S 4). Two 
sub-fields were stacked to simulate a field with $\mu_B = 
22.0$ mag arcsec$^{-2}$, while three were combined to simulate 
$\mu_B = 21.6$ mag arcsec$^{-2}$. All four sub-fields were combined to simulate 
$\mu_B = 21.3$ mag arcsec$^{-2}$.

	The magnitudes of stars in the stacked images were measured 
using the procedures discussed in \S 3.1, and the resulting CMDs are shown 
in Figures 2 and 3. The left hand column of each figure shows 
the CMDs from the unstacked data, while the right hand column shows the CMDs 
of the stacked images. The simulated surface brightness of each stacked 
field is also listed, along with the corresponding major axis radius from 
the Walterbos \& Kennicutt (1987) light profile.

	As expected, crowding elevates the faint limit of the CMDs 
as one moves to progressively higher surface brightnesses. An interesting 
practical result is that the photometry of sources with $g' < 23$ and 
$r' < 22$ appears not to be sensitive to surface brightness in the density regime 
investigated here. The brightest main sequence stars and AGB 
stars in these data can thus be used to trace the distribution of young stars to 
within a few kpc of the galaxy center.

\section {COLOR-MAGNITUDE DIAGRAMS}

	We define six regions in the M31 disk to facilitate 
analysis and subsequent discussion. Two regions are associated with 
the rings of UV and FIR emission at R$_{GC} = 10$ and 14 kpc,  
and these will be refered to as Ring 10 and Ring 14. The Inner Disk is 
defined to be the region that is interior to Ring 10. 
The image stacking experiments in \S 3.4 indicate that crowding is 
an issue for all but the brightest stars with R$_{GC} \sim 2$ kpc, and so 
the minimum radius for the Inner Disk is set at R$_{GC} = 3$ kpc. The 
Inner Disk thus probes a region where the disk light dominates over 
that of the bulge (e.g. Figure A1 of Tempel et al. 2011). 
The Middle Disk is located between Ring 10 and 14, while the Outer Disk is 
external to Ring 14.

	The Outer Disk is split into two radial intervals. 
This is done in part because the outer radius of the M31 disk is not clearly 
defined. In addition, there is evidence that the properties of 
star formation in low density environments may differ from 
those in the inner regions of disks (Bigiel et al. 2010), while secular processes 
may play a key role in defining the stellar content at large radii (e.g. Roskar et 
al. 2008; Sanchez-Blazquez et al. 2009). The outermost regions of disks may then 
be an environment that is distinct from the inner disk, containing 
population gradients with physical drivers that differ from those at smaller radii.

\subsection{$(g', u'-g')$ CMDs}

	The $(g', u'-g')$ CMDs of roughly 2.1 million sources 
in various radial intervals are shown in Figure 4. 
Hess diagrams are shown in Figure 5 to allow the relative properties 
of objects in the faint portions of the CMDs to be examined.
The radial distances listed in each panel of Figures 4 and 5 
are in the plane of the disk and assume 75$^o$ inclination. This inclination 
produces roughly circular rings when observations of warm dust emission
are de-projected to simulate the face-on appearance of M31 (Gordon et al. 
2006). While the assumption of a single inclination is an 
obvious pragmatic choice, it is a simplification 
given that the gas disk of M31 is warped (e.g. Corbelli et al. 2010).
This warping, coupled with the orientation of M31 on the sky, 
frustrates efforts to extract clean radial samples that are free of 
annulus-to-annulus contamination. 

	Young main sequence stars form a prominent plume in the CMDs, 
with $u'-g' \sim 0$ at the bright end, and $u'-g' \sim 0.5$ near the faint end. 
Bright blue stars are not restricted to Ring 10 
and Ring 14, and are seen in all CMDs with R$_{GC} < 21$ kpc. 
Recent star formation has thus not been restricted exclusively to Ring 10 and 
Ring 14. Many of the blue objects with $g' < 18$ (M$_{g'} \leq -7$) are bright blue 
supergiants (BSGs), although some of these may be 
young compact star clusters. The diffuse sequence of objects that is displaced 
$\sim 0.6$ magnitude in $u'-g'$ redward of the 
main sequence, and is most pronounced in the Inner Disk, Ring 10, 
and Middle Disk CMDs with $g'$ between 20 and 22, is 
due to Helium burning stars that are at the blue extrema of their evolution. 
This same feature is seen in Figure 2 of McQuinn et al. (2011).

	While the primary probe of evolved stars in our data is the $(r', g'-r')$ 
CMD (\S 4.2), evolved red stars are also present in 
the $(g', u'-g')$ CMDs. A population of red objects 
with $u'-g > 1$ and $g' > 23$ is seen in the Hess diagram of the Inner 
Disk region in Figure 5. Isochrones (see below) indicate that evolved stars with 
ages of $\sim 0.2$ Gyr pass through this part of 
the $(g', u'-g')$ CMD. That these red sources occur in large numbers in 
the Inner Disk $(g'- u'-g')$ CMD when compared with the Ring 10 and the Middle 
Disk CMDs suggests that the SFR in the Inner Disk was elevated with respect 
to regions of M31 at larger radii within the past 200 Myr (\S 6). 

	Background galaxies contribute significantly to the source counts near the 
faint limit of these data. The majority of background objects have red colors, and 
populate the jumble of red sources with $g' > 23$ that is most obvious in the 
Outer Disk CMDs. Not all background objects have red colors, and Davidge \& 
Puzia (2011) found that a significant fraction of the objects with blue colors at 
large radii in M33 are star-forming regions in moderately 
distant background galaxies. This being said, source counts in areas where the 
number of M31 disk stars is expected to be small indicate that many of 
the blue objects with $g' > 20$ in the Outer Disk 2 CMD are {\it bona fide} 
main sequence stars in M31.

	M31 is viewed at intermediate Galactic latitudes, and foreground stars 
dominate the diffuse spray of objects with $g' < 22$ and $u'-g'$ between 1 and 3 
that grows more prominent towards larger R$_{GC}$. The 
blue edge of this sequence is defined by the MSTO 
of stars in the outer disk and halo of the Galaxy, while the red envelope reflects 
the properties of low mass main sequence stars. Not 
all of the sources with $u'-g'$ between 1 and 3 are in the foreground, and 
BSGs with $u'-g' < 2$ and $g' < 22$ (i.e. to the immediate right of the M31 
main sequence) are evident in the CMDs of sources with R$_{GC} < 15$ kpc.

	The $(g', u'-g')$ CMDs of sources in Ring 10 and Ring 14 are compared with 
solar metallicity isochrones from Girardi et al. (2004) in Figure 6. 
Metallicity gradients are present in many spiral galaxies, and so a comparison 
with isochrones having only one metallicity may seem suspect. However, 
the metallicities of BSGs in M31 with R$_{GC}$ between 
5 and 12 kpc do not vary with radius
(Trundle et al. 2002), suggesting that the interstellar 
medium of M31 is well mixed, and that the metallicity gradient 
among moderately young stars is modest. Therefore, with the exception of 
a single Z = 0.004 sequence to demonstrate metallicity effects, the comparisons 
with isochrones are restricted to a single metallicity. 

	The image stacking experiments discussed in \S 3 indicate that crowding 
does not affect greatly the photometric properties of objects with M$_g' < -2$, 
the majority of which are main sequence stars with ages 
$< 100$ Myr. It is thus significant that the zero age 
main sequence (ZAMS) defined by the isochrones falls along the blue envelope 
of main sequence stars, while the terminal age main sequence (TAMS), 
defined by the red extent of main sequence evolution, follows 
the red envelope of the blue plume. That the observed blue plume is bracketed by 
these two evolutionary phases suggests that age is the dominant driver 
of main sequence width in our data, rather than photometric errors or 
differential reddening.

	Ring 10 and Ring 14 both contain very young stars. The brightest blue 
objects in Ring 10 and Ring 14 fall more-or-less along the 5 Myr 
isochrone in Figure 6. This is not unexpected given that 
these annuli are known to contain areas of very recent 
star formation. It is also evident from the isochrones that the part of the 
CMD with M$_{g'} > -2$ and $u'-g' > 1$ contains evolved stars with ages 
$\geq 0.2$ Gyr. Hence, the red concentration of objects near the faint limit 
of the $(g', u'-g')$ CMD contains intermediate age stars.

	The blue core helium-burning (BHeB) sequence in M31 
is not well-matched by the locus of blue loops in the Z = 0.019 isochrones, 
in the sense that the locus of blue loops during supergiant evolution 
is almost 1 magnitude in $u'-g'$ redward of the observed 
sequence. McQuinn et al. (2011) also found differences between the observed and 
predicted colors of the BHeB sequence in their sample of dwarf galaxies. Such 
disagreements with observations may be due to -- for example -- errors in 
the physics used to generate the models, uncertainies 
in circumstellar extinction, or uncertainties in the 
metallicities adopted for the stars. An interesting result from 
Figure 4 of McQuinn et al. (2011) is that the colors predicted for BHeB stars do 
not change greatly with metallicity when Z $< 0.019$. 
The blue loop of the Z = 0.004 isochrone in Figure 6 overlaps the M31 BHeB 
sequence, and one {\it possible} explanation for the poor 
agreement between the observed and predicted color of the BHeB sequence in Figure 6 
is that the stars in M31 have a metallicity that is no more than 
one half solar (i.e. [M/H] $\sim -0.3$). However, we reiterate that metallicity is 
just one of the factors that may affect the color predicted for the BHeB sequence.

	The comparisons in Figure 6 assume the same reddening 
for each star. Reddening maps have been constructed for M31 
using the distribution of warm and cool dust, and these could be used to 
estimate reddening for individual sources. However, our MegaCam data are not 
consistent with the radial extinction trends predicted 
from dust emission. This is demonstrated in Table 2, where the mean $u'-g'$ 
color, $\overline{u'-g'}$, of sources with $g'$ between 
21.75 and 22.25 is given for 5 radial intervals. 
An iterative $2.5\sigma$ rejection filter was applied to suppress outliers. 
The standard deviation about the mean, $\sigma_{u'-g'}$, is also shown, along with 
the mean optical depth at each R$_{GC}$, $\tau_{g'}$, taken from 
Figure 4 of Tempel et al. (2011).

	The entries in Table 2 indicate that the plume 
containing main sequence stars and BSGs has a dispersion 
$\pm 0.15$ magnitudes at $g' = 22$, supporting the notion that photometric errors 
make only a modest contribution to the width of the main sequence.
The entries in Table 2 further indicate that neither the mean color of the main 
sequence nor the dispersion about the mean changes significantly with 
radius between R$_{GC} = 5$ kpc and 15 kpc. Indeed, the mean 
$u'-g'$ color of the main sequence in this large range of R$_{GC}$ varies by 
only a few millimagnitude. To be sure, there are area-to-area variations in 
extinction within each annulus. However, the radial consistency of the mean 
$u'-g'$ color indicates that these variations average out 
to a remarkable degree when stellar measurements are combined azimuthally.

	The uniform color of the main sequence ridgeline in Table 2
is contrary to what might be expected based on the optical depth measurements from 
Tempel et al. (2011), which predict systematic radial variations in total 
extinction, with $\tau_{g'}$ peaking in Ring 10. Such an inconsistency would occur 
if the radial $\tau_{g'}$ measurements obtained from dust emission do not 
apply to the majority of stars in each angular resolution element.
Indeed, a core assumption made by Tempel et al. (2011) is that 
cool dust is uniformly distributed. There almost certainly $is$ a uniformly 
distributed cool dust component throughout the M31 disk; however, there are also 
areas of cool dust emission that are concentrated on scales that are smaller than 
the angular resolution of the facilities that have probed FIR emission, and such 
a component will skew $\tau$ measurements. 

	The Tempel et al. (2011) optical depth measurements are 
based on data from the Spitzer telescope, the angular resolution of which is 
roughly 40 arcsec, or $\sim 0.15$ kpc in M31, at $160\mu$m. Each resolution element 
then has a spatial extent that is roughly one order of magnitude larger than 
that of a typical giant molecular cloud (GMC). If a significant part of the cool 
dust emission within a resolution element originates from GMCs then 
the extinction of objects that are exterior to the GMCs, but still within the 
same observed angular resolution element, will be over-estimated. The stars that 
are obscured in GMCs by the cool dust may even be so heavily extincted that they 
are not detected at visible wavelengths, and if this is the case then the dust 
and observed stars are spatially de-coupled. In summary, if there 
is a cool dust component in M31 that is concentrated on GMC-like spatial scales, 
then the stars within at least some Spitzer resolution elements will not 
be subject to the level of dust extinction that is predicted from dust emission. 

\subsection{$(r', g'-r')$ CMDs}

	The $(r', g'-r')$ CMDs of roughly 3.6 million sources in the M31 disk are 
shown in Figure 7, while the corresponding Hess diagrams are shown in Figure 
8. Foreground stars form a spray of objects at the bright end 
of the CMDs in Figure 7 that has a well-defined blue and red cut-off; 
this sequence falls between $g'-r' = 0$ and 2, and is most obvious when $r' < 21$. 
There is also a prominent blue plume with $g'-r' \sim -0.1$ that contains main 
sequence stars. The blue plume is less well-defined in the $(r', g'-r')$ CMDs than 
in the $(g', u'-g')$ CMDs, and it is demonstrated below that this is probably 
a consequence of the trajectories followed by stars on this CMD as they 
evolve off of the ZAMS. 

	A collection of objects with $r' > 22$ and $g'-r'$ between 0 and 2 forms 
a prominent feature in the $(r', g'-r')$ CMDs. At small and intermediate R$_{GC}$ 
this part of the CMD is dominated by a mix of red core helium-burning (RHeB) 
giants/supergiants and stars that are evolving on the AGB. 
Background galaxies make a progressively larger contribution to the faint 
red end of the CMDs as one moves to larger R$_{GC}$.

	The $(r', g'-r')$ CMDs of Ring 10 and Ring 14 are compared with solar 
metallicity isochrones from Girardi et al. (2004) in Figure 9. 
The isochrones follow a near-vertical trajectory on the 
$(r', g'-r')$ CMD after leaving the ZAMS; this continues past the TAMS 
until the onset of red loops, smearing the main sequence in a vertical 
direction on the CMDs, and making the blue plume 
in the $(r', g'-r')$ CMD near $g'-r' \sim -0.1$ more diffuse than in the 
$(g', u'-g')$ CMD. It is encouraging that the blue plume 
in the $(r', g'-r')$ CMDs is bracketed by the ZAMS 
and the inflexion points in the isochrones that mark the onset of red loops. 

	The isochrones indicate that the red jumble is made 
up of stars in a mix of evolutionary stages, some of which are RHeB and others 
evolving on the AGB. Given that the timescale for core Helium burning is 
much longer than the timescale to evolve on the AGB then stars in the former 
stage of evolution are the dominant population. The 160 and 240 Myr isochrones pass 
through the main concentration of red sources that have M$_{r'} > -3$, indicating 
that the majority of these stars have progenitor masses $\sim  3 - 4$M$_{\odot}$.

	Line blanketing affects the photometric properties 
of very red evolved stars at visible and red wavelengths, causing 
the most luminous portions of the AGB at moderate and higher metallicities 
to bow over on the $(r', g'-r')$  CMDs (e.g. Bica et al. 1991). This is evident 
in Figure 9, as the 160 Myr isochrone plunges to fainter M$_{r'}$ when 
$(g'-r') > 1.5$. The most luminous portions of the AGB are almost certainly 
missing from these CMDs, and so identifying the AGB-tip is problematic.

\subsection{The $g'-r'$ Color Distribution}

	The color distributions of stars can be used to investigate 
field-to-field variations in the recent SFH in a purely empirical manner. The 
$g'-r'$ histogram distributions of sources in three R$_{GC}$ intervals 
with M$_{r'}$ between --4 and --3, which corresponds to $r'$ between 20.9 
and 21.9 based on the apparent $r'$ distance modulus, are compared 
in Figure 10. Comparisons with isochrones in Figure 9 indicate that 
the blue peak is populated by stars with ages $< 100$ Myr, while the 
red peak contains stars with ages between 60 and 200 Myr. 

	The number counts in each radial interval have been corrected statistically 
for foreground and background contamination using source counts in areas 
that are external to Outer Disk 2. These control fields, the location of 
which are indicated in Figure 1, were examined by eye to ensure that 
they do not contain obvious concentrations of young stars. The control 
fields almost certainly contain some intermediate age stars that belong to M31, and 
so the number counts in these areas may (slightly) overestimate 
the level of foreground and background contamination at the faint end.

	That Ring 10 has been an area of star-forming activity during the 
past 100 Myr is clearly evident from the number of blue 
stars in the top panel of Figure 10. The mix of young 
stars (i.e. those in the blue color peak) to intermediate age stars 
(i.e. those in the red color peak) changes with radius, and these variations are 
explored in the lower panel of Figure 10, where the color distributions 
have been normalized according to the number of objects 
with $g'-r'$ between --0.25 and 0.25. There 
is a wide dispersion in the relative amplitudes of the blue and red peaks. 
The ratio of red to blue objects in the Inner Disk is substantially larger 
than in either Ring 10 or the Middle Disk, indicating that the SFH in this part of 
M31 was skewed to higher levels of activity 60 -- 200 Myr in the past. This is 
another manifestation of the pronounced red stellar concentrations that are seen 
in the Hess diagrams of the Inner Disk in Figures 5 and 8. The 
SFH of the Inner Disk during intermediate epochs is re-visited in \S 6.

	The blue sequence in the Middle Disk peaks at a bluer $g'-r'$ 
color than at smaller radii. A comparatively low reddening for the Middle Disk 
is probably not driving the peak color of blue stars, as there is not a 
corresponding shift in the color of the main sequence on the $(g', u'-g')$ 
CMD (\S 4.1). Rather, the trajectory of the isochrones on 
the $(r', g'-r')$ plane makes the $g'-r'$ color distribution of blue stars 
more susceptible to variations in the recent SFH. 
The blue sequence will narrow as stars at the older end of the $\sim 
100$ Myr age range are removed, as these stars have redder colors than 
younger stars in a given magnitude interval. This may 
not be accompanied by a skewing of the color distribution of red stars, as 
these objects span a much larger range of ages than main sequence stars 
with the same M$_{r'}$, and so their mean colors are less susceptible to 
variations in the SFH.

\section{MAIN SEQUENCE STARS AND THE RECENT STAR-FORMING HISTORY}

\subsection{Main Sequence LFs}

	As in our study of M33 (Davidge \& Puzia 2011), we use the $u'$ 
LF of blue stars to probe the SFH of the M31 disk. 
Our $u'$ photometry includes main sequence stars 
throughout much of the M31 disk with ages $\sim 100  - 200$ Myr, 
allowing the SFH in this timeframe to be directly probed. The damping time 
for large scale star-forming activity in disks is $10^8 - 10^9$ years 
(e.g. Leitherer 2001), and so elevated SFRs may linger for this time after 
they are initiated. Therefore, inferences can be made into the more 
distant past than would otherwise be permitted from resolved stars.

	Following Davidge \& Puzia (2011), LFs were constructed of sources with 
$(u'-g')_0$ between --0.5 and 0.5. The majority of objects in this color range 
are main sequence stars. While most foreground and background 
objects in the CMDs have red colors, there are modest numbers of contaminants that 
fall within the color interval considered here. These include blue horizontal 
branch stars in the Galactic halo, and star-forming regions in moderately 
distant spiral galaxies. We account for these statistically using number counts 
in portions of our MegaCam fields that are offset from the disk of M31 (\S 4.3).
The mean LF in these control regions follows a power-law, and a 
characteristic exponent and zeropoint were determined by using the method of least 
squares to fit a power-law to this LF. This fitted relation 
was then subtracted from each of the observed LFs, 
after scaling to account for differences in angular coverage. 

	The LFs of various radial intervals are compared in Figure 11. The 
number counts in the top panel are as observed on the sky, and these are given 
in units of counts arcmin$^{-2}$ per 0.5 magnitude interval in M$_{u'}$. The 
$\sim 1$ dex offset between the Middle and Outer Disk LFs is consistent with 
the Walterbos \& Kennicutt (1987) surface brightness profile, as is the 0.5 dex 
(i.e. 1.25 magnitude arcsec$^{-2}$) dispersion between the LFs with R$_{GC}$ 
in the range 5 - 15 kpc. This broad consistency between star counts and surface 
photometry is encouraging, and suggests that the bright blue stars detected 
in the MegaCam data are not affected by crowding.

	The LFs in the bottom panel of Figure 11 have been normalized to match 
the Ring 10 LF between M$_{u'} = -3.75$ and --5.25, thereby allowing 
the shapes of the LFs to be compared. The LFs follow power-laws, and 
the Outer Disk 2 LF is steeper than the rest. The mix of massive and intermediate 
mass stars in the Outer Disk thus differs from that in the Inner Disk.

	The LFs of the Inner Disk, Ring 10, and the Middle Disk in the lower 
panel of Figure 11 are similar. This is perhaps surprising given the 
concentration of star-forming pockets in Ring 10 (\S 6). However, it should be 
kept in mind that the LFs are azimuthal averages, and isolated star-forming 
pockets will only influence the mean LF if they occur in large numbers. In fact, 
while Ring 10 and Ring 14 contain areas of intense recent star formation, 
they do not stand out in the UV color profile measured by Thilker et al. (2005); 
rather, the UV color stays roughly constant between 10 and 15 kpc. The comparisons 
in Figure 11 indicate that the mix of young to intermediate age stars in Ring 10 
is not different from that in the Inner or Middle Disks -- when averaged over large 
areas, the SFHs of these regions have been broadly similar during the past 
100 -- 200 Myr. 

\subsection{A Comparison with M33}

	A comparison of the properties of bright main sequence stars in M31 and M33 
provides a purely empirical means of assessing the relative star-forming activities 
of these galaxies during recent epochs. To be sure, 
star-forming activity in disk galaxies depends on a number 
of factors, such as environment and mass, and it would be of interest to 
compare the stellar content of M31 with that of other large galaxies of similar 
morphological type; however, such datasets do not yet 
exist. Differences in size and morphological type aside, M33 is an 
important comparison object as it is at roughly the same distance of M31, and has 
been the subject of a number of investigations. The SFHs of M31 and M33 have 
probably also been coupled at some point in the past (\S 1.3). Finally, the 
specific SFR (sSFR) of M33 is near the mid-point of spiral galaxies in general 
(Munoz-Mateos et al. 2007), making it a benchmark for activity in a 
`typical' star-forming galaxy. The M33 MegaCam dataset examined by Davidge \& Puzia 
(2011) was recorded during observing conditions that were similar to those of the 
present M31 observations, and are adopted here for comparisons with M31.

	Main sequence stars with ages $< 10$ Myr are 
traced throughout much of M33 (Davidge \& Puzia 2011) and M31 (\S
3), and these stars provide a direct means of comparing 
the SFRs of these galaxies. Davidge \& Puzia (2011) found 
3000 massive main sequence stars and BSGs in M33 with ages $\leq 10$ 
Myr. Applying the M33 selection criteria of M$_{u'} \leq -5.8$ to the current data, 
we find that there are $10^3$ similar objects in M31. The SFR in M31 during the 
past 10 Myr has thus been one third that of M33. For comparison, the 
integrated H$\alpha$ luminosities of M31 and M33 are comparable (e.g. Kennicutt 
et al. 2008), while M31 has a slightly higher FIR flux than M33 (Rice et al. 1988). 

	The sSFR is a measure of the rate at which stellar mass grows due to
star formation. The bulk of the $K-$band light from most galaxies has its origins 
in old and intermediate age stars, and so integrated $K-$band brightness 
serves as a crude proxy for total stellar mass. Adopting the apparent $K$ 
magnitudes of M31 and M33 from Jarrett et al. (2003) and a distance 
modulus of 24.92 for M33 (Bonanos et al. 2006), then the integrated total 
brightnesses of these galaxies are M$_K = -23.4$ (M31) and 
M$_K = -20.9$ (M33). Based on the numbers of blue stars with ages $\leq 10$ Myr 
(see above) and their integrated $K-$band brightnesses, then the sSFR of M31 is 
only $\sim 3\%$ that of M33. The difference between the sSFRs of M31 and M33 is 
almost certainly greater than this, as it has been 
assumed that M31 and M33 have identical $K-$band mass-to-light ratios (M/L$_K$). In 
actual fact, given the higher relative fraction of young stars in M33, then 
the M/L$_K$ of M33 is probably lower than that of M31. M/L ratios 
at a given age depend on metallicity, although for effective ages 
$\geq 10^9$ years the differences between the M/L ratios 
predicted by Z = 0.02 and Z = 0.008 models are miniscule 
(e.g. Figure 4 of Mouhcine \& Lancon 2003). Minor differences in the relative 
M/L ratios of these galaxies notwithstanding, it is clear that M31 is practically 
dead in terms of recent star-forming activity when compared with M33.

	The M31 and M33 LFs, normalized in the interval M$_{u'} =$ --4 to --5, 
are compared in Figure 12. The M33 LF is that of objects with R$_{GC}$ between 
6 and 8 kpc from Davidge \& Puzia (2011). The LF of sources in 
this part of the galaxy is representative of the main body of the M33 disk. 

	There are conspicuous differences between the M31 and M33 LFs in Figure 12. 
These differences are likely not a result of differences in metallicity between 
stars in the disks of M31 and M33. While model LFs of blue sources with Z = 
0.019 and Z = 0.008 do differ, the differences are largely restricted to 
M$_{u'} < -6$. This magnitude regime is dominated by BSGs, the properties 
of which are affected by metallicity-dependent mass loss rates. When M$_{u'} 
> -6$ main sequence stars dominate, and the model LFs of blue sources show only 
slight differences.

	When compared with the M33 disk, Outer Disk 1 has a lower number of stars
with ages of a few tens of Myr with respect to stars that formed a few hundred Myr 
in the past. When compared with the M31 LFs, the M33 LF is also deficient in stars 
with M$_{u'}$ between --2 and --4, which is a M$_{u'}$ range that corresponds to 
MSTO ages of 40 -- 100 Myr. Davidge \& Puzia (2011) found that 
the LF of bright main sequence stars that formed during the past few hundred Myr
in M33 is well matched by a constant SFR (cSFR) model, and
the comparisons in Figure 12 indicate that such a SFH does not hold for M31. 
The SFH of the M31 disk is examined in more detail in the next section. 

\subsection{Comparisons with Models}

	We use a forward modelling approach, in which
model LFs are constructed for pre-defined SFHs and are then compared 
with the observations. Models were generated from the Z = 0.019 Girardi et al. 
(2004) isochrones using routines in the STARFISH package (Harris \& Zaritsky 2001). 
The models assume an IMF power-law index $\alpha = 
-2.7$, which Kroupa et al. (2003) find holds for massive stars. 
Models with a Salpeter (1955) IMF differ only slightly 
from those used here (e.g. Figure 11 of Davidge \& Puzia 2011).

	The $u'$ LF of main sequence stars provides a relatively robust means of 
investigating SFHs, as the main sequence is probably the best understood phase of 
stellar evolution. Still, there is an incomplete understanding of the physical 
processes that affect the structure and evolution of massive main sequence 
stars. These include uncertainties in convection, which 
is the dominant mode of energy transport in the central 
regions of massive stars, mass loss, and rotation. While these uncertainties 
affect the post-ZAMS models of massive stars, it is encouraging 
that the ZAMS and TAMS loci predicted by the isochrones match the blue 
and red envelopes of the main sequence on the CMDs (\S 4).

	Statistical studies of main sequence stars are susceptible to uncertainties 
in the properties of close companions, which are not 
resolved at the distance of M31. These uncertainties are not an issue 
for statistical studies of evolved stars, as the light from 
the most evolved star in a close stellar system swamps 
that from less evolved companions at visible wavelengths. However, if all the stars 
in a close system are on the main sequence then -- depending on the mass 
ratio -- companions may contribute a significant fraction of the 
total system light. While many of the sources in our data are 
unresolved systems, if the statistical properties of these 
systems (e.g. the frequency of binary or higher-order systems, the mass 
ratio of components, etc) do not change with the mass 
of the primary then the main sequence LF will still yield useful 
information for a differential examination of the SFH.

	The MegaCam data sample stars in a wide range of evolutionary states, 
all of which provide information about the SFH of M31. However, combining 
information from different evolutionary states to investigate the SFH may lead to 
complications when interpreting the results, as the dominant sources of 
uncertainties in the input model physics vary with evolutionary 
phase. Comparisons of the SFHs of galaxies at different distances may be 
most prone to uncertainties related to evolutionary state, 
as the SFHs may be based on objects in very different stages of evolution. 
Gogarten et al. (2010) suggest that this may be a factor when comparing the 
SFHs of M33 and NGC 300. The impact of uncertainties in models of the 
most advanced stages of stellar evolution on SFHs has been investigated by 
Melbourne et al. (2010), who compute separate SFHs for a dwarf galaxy using main 
sequence stars and AGB stars. The two SFHs differ substantially, and Melbourne et 
al. (2010) attribute this difference to an incomplete empirical calibration of the 
AGB models, which rely heavily on the LMC and SMC as benchmarks. 

	The models used here assume a fixed metallicity, and so do not account for 
metallicity evolution. While the mean metallicity of a disk grows with time as 
material that is processed in stars is recycled into the ISM, a fixed metallicity 
is appropriate given that our models cover a time span that is short when 
compared with the enrichment timescale of a disk. 
Indeed, the age-metallicity relation of the solar neighborhood 
during recent epochs has a shallow slope (e.g. Edvardsson et al. 1993), and the 
rate of metallicity evolution in M31 will be even more 
subdued than this given the lower SFR.

	The most basic SFH model is that of a simple stellar population 
(SSP), in which all stars have the same age and metallicity; SSP 
LFs are also the building blocks from which LFs that track more 
elaborate SFHs are constructed. Because galaxies are 
composite stellar systems, it is unlikely that their LFs will 
be well matched by SSP models. Still, if a galaxy experiences a single large 
episode of star formation that is caught early in its evolution then 
the LF of the brightest stars may follow that of an SSP. 

	The $u'$ LFs of main sequence stars in three radial intervals are compared 
with a log(t$_{yr}) =6.8$ SSP model LF in Figure 13. The Ring 10 LF is similar 
to that of the Inner and Middle Disk regions (Figure 11), and so comparisons with 
models are restricted to the four intervals shown in Figure 13. 
M31 is a composite stellar system, and so it is not 
surprising that there is poor agreement between the model SSP LF and the 
observations. Still, the Ring 10 LF, which samples a region with a high 
concentration of recent star formation, has an exponent that comes closest to 
matching that of the SSP model.

	The SFHs of isolated, passively evolving disks may be affected by 
stochastic effects, with spikes in star-forming activity 
due to events such as the passage of spiral arms and the propagation of 
density waves from SNe in young clusters and associations. 
Evidence for variations in the SFH of the Solar neighborhood due to the passage 
of spiral arms has been found by Hernandez et al. 
(2000). Nevertheless, the impact of stochastic events will average out 
when considered over long timespans and large areas, and the SFHs of passively 
evolving disks might be expected to come close to that of a cSFR. 

	The $u'$ LFs are compared with cSFR models in Figure 14, and there is 
much better agreement with the observations than in Figure 13.
In fact, the cSFR model provides a reasonable match to the Outer Disk 2 
LF, and is a fair representation of the Outer Disk 1 LF. Still, with the 
exception of the Outer Disk 2 LF, the cSFR model LFs in 
Figure 14 tend to be steeper than the observed LFs. Indeed, while 
the cSFR model is a reasonable match to the Outer Disk 1 
observations with M$_{u'} > -5$, which corresponds to MSTO ages $\geq 10$ Myr, 
it underestimates the number of stars at the bright end. These differences 
are even more evident when the Ring 10 and Ring 14 LFs are compared with the cSFR 
model. 

	Given the differences between the observations and models in Figure 14, 
we investigated SFHs in which the SFR during the past 10 Myr has 
increased with respect to that expected for a cSFR at earlier epochs.
Models were generated in which the recent SFR was increased by 
various amounts, and the results were compared with the observed LFs; 
the best matches, selected by eye, are shown in Figure 15. Also shown in each 
panel is b$_{100}$, which is the ratio of the number of stars that formed within 
the past 10 Myr to the number that formed in the interval 10 -- 100 Myr. The Outer 
Disk 2 LF is not shown in Figure 15, as it is represented adequately by the cSFR 
model.

	Models with elevated recent star-forming activity better match 
the LFs of the three regions shown in Figure 15 than the cSFR models.
Higher recent SFRs are required to match the Ring 10 and 
Ring 14 LFs than is required to match the Outer Disk 1 LF. While this is 
consistent with Ring 10 and 14 being areas of comparatively intense recent 
star-forming activity, if elevated levels of recent star formation occur 
in only part of Outer Disk 1 (i.e. at the smallest R$_{GC}$ in this region) 
then the effect of any recent increase in the SFR will 
be diluted. The differences in the amplitude of 
recent star-forming activity aside, it is clear that the recent 
upswing in star-forming activity has not been restricted to Ring 10 and 14, but 
occured over a large fraction of the disk -- the change in the recent SFR is not 
a localized phenomenon that is restricted to only a small part of the M31 disk, 
but is more global in nature.

\section{THE SPATIAL DISTRIBUTION OF STARS IN THE M31 DISK}

\subsection{Young Main Sequence Stars}

	The spatial distribution of stars in a known age range is an important 
probe of galaxy evolution. Samples of young main sequence stars (YMS; ages 
$< 10$ Myr) and intermediate-age main sequence stars (IMS; ages $\sim 100$ Myr) 
were selected from the MegaCam data based on their location in the $(g', u'-g')$ 
CMD. The boundaries of the YMS and IMS regions in Figure 16 were defined using 
isochrones as guides, while also keeping in mind the results of the artificial 
star and stacking experiments discussed in \S 3. 

	The distributions of objects in the YMS and IMS samples are shown in 
Figures 17 (as observed on-sky) and 18 (as would be observed face-on). The face-on 
distributions assume that the stars are in an infinitely thin, unwarped disk, and 
departures from these assumptions will cause blurring of structures in the 
de-projected distributions. Some blurring is expected, as the gas disk of 
M31 is warped (Corbelli et al. 2010). 

	It is evident from Figures 17 and 18 that (1) recent star-forming 
activity in M31 has not been uniformly distributed with azimuthal angle, and (2) 
knots of stars in both samples are seen throughout the disk. The stars 
in the YMS sample tend to be tightly clustered and define partial arcs. 
The distribution of YMS stars indicates that the most intense 
area of recent star formation was in the north east quadrant of the disk. 

	There is broad agreement between the YMS distribution and images 
of UV (Figure 1 of Thilker et al. 2005) and MIR 
(Figure 1 of Gordon et al. 2006) emission. While peaks in HI emission shown in 
Figure 1 of Unwin (1983) coincide with areas of high 
YMS density in Ring 10, at radii $< 47$ arcmin (R$_{GC} < 10$ kpc) there are 
collections of YMS stars with no corresponding HI feature. In contrast, there is 
much better agreement between the location of YMS concentrations and the peaks in 
CO (1--0) emission mapped by Nieten et al. (2006).

	Stars in the YMS sample are seen at moderately small R$_{GC}$. 
Large-scale, well-defined structures that contain bright blue stars
are seen near 20 arcmin (i.e. R$_{GC} = 4$ kpc), while individual objects 
in the YMS sample are detected to within 10 arcmin (R$_{GC} \sim 2$ kpc) of 
the nucleus. This R$_{GC}$ is where the bulge and disk contribute comparable 
amounts of light at visible wavelengths (e.g. Tempel et al. 2011), and is where 
the FUV--NUV color in the main body of the galaxy is reddest (Thilker et al. 2005). 

	Ring structures are much more pronounced in the IMS distribution 
than in the YMS distribution. The IMS sample contains stars that 
formed over a wider range of ages than the YMS stars, and so the effects of 
spatial fluctuations in star-forming activity are expected to average out. 
The more-or-less uniform azimuthal distribution of 
IMS objects in the rings suggests that the location of star 
formation in the immediate past differed from the present day, and that areas of 
active star formation in the rings have moved through a wide range of 
position angles within M31.

	That Ring 10 is seen in samples of objects that are as old as 100 Myr 
indicates that the event that spurred its formation must have occured at least 
100 Myr ago. In fact, it is somewhat surprising that rings and arcs are 
present at all in the IMS sample, given the rate at which stars 
in the disks of nearby star-forming galaxies acquire random velocities 
with time as they interact with giant molecular clouds 
(GMCs). The velocities acquired in this manner can cause 
disk stars to move substantial distances over timespans of 100 Myr. For example, 
spiral structure in M33 is well-defined by stars with ages $\leq 10$ Myr, but 
is not apparent in the distribution of stars with ages of 100 Myr (Davidge \& 
Puzia 2011). The presence of organized structures in the disk of M31 that are 
defined by objects in the IMS sample suggests that stars in these parts of M31 do 
not attain the same random motions as their counterparts in M33. The narrow 
distribution of orbital angular momentum that is expected for star-forming material 
in rings undoubtedly contributes to restricting the time evolution of 
random velocities among stars in these structures. 

\subsection{Evolved Red Stars and Age Gradients}

	A sample of RHeB$+$AGB stars has been selected 
using the boundaries indicated on the $(r', g'-r')$ CMD in Figure 16. 
The objects in this part of the CMD formed 60 -- 200 Myr 
in the past, and so probe earlier epochs than the IMS sample. The faint 
limit of the extraction region in Figure 16 reflects the results of the data 
characterization experiments discussed in \S 3.

	The face-on distribution of objects in the RHeB$+$AGB sample is shown in 
Figure 19. The distribution observed on the sky is not shown, as the density 
of red sources makes it difficult to detect structure because of the 
orientation of M31 on the sky. As in Figure 18, the major axis of M31 points 
upwards. The concentration near (dX,dY) = (--60,--20) is centered on M32; the dense 
central regions of M32 -- where stars are not resolved -- appears as a 
hole in the RHeB$+$AGB distribution. The concentration near (dX,dY) = (100, 0) 
is probably associated with NGC 205 (see Figure 1). 

	The distribution of RHeB$+$AGB stars in Figure 19 is asymmetric 
about the horizontal axis, in the sense that the stellar density 
above and to the left of the galaxy center (i.e. along the north east arm of the 
semi-major axis) is higher than that below the galaxy center. 
There are also asymmetries to the left and right of the galaxy center, which fall 
along the minor axis. However, the orientation of M31 on the sky is 
such that the distribution of objects along the minor axis is more prone to 
uncertainties arising from -- for example -- the inherent thickness of the disk 
and the presence of stars in the bulge of M31, both of which may affect the 
de-projected stellar distribution. We thus focus on the asymmetry along the 
major axis, where the de-projected distribution is most secure.

	Number counts in an 8 arcmin wide strip along the 
north east and south west segments of the major axis are shown in Figure 20. 
The density of RHeB$+$AGB stars along the NE axis within 50 arcmin ($\sim 10$ 
kpc) of the galaxy center is higher than along the SW axis. 
At $\Delta Y = 25$ arcmin these stars have a density that is 0.4 dex higher on the 
NE axis than at the same point on the SW axis. This suggests 
that the north east part of the galaxy was an area of more 
intense star-forming activity $100 +$ Myr in the past than the south west part. 
Persistent large scale concentrations of elevated star-forming 
activity are seen in other nearby disk galaxies, such as NGC 253 (Davidge 2010). 

	The number counts of objects in the IMS sample are also shown in Figure 20, 
and it is evident that the IMS and RHeB$+$AGB samples have very different radial 
distributions. Star-forming rings form a 5 kpc periodicity in the 
distribution of IMS stars, and these structures are most pronounced along the NE 
axis. The rings have an amplitude of $\sim 1$ dex in the IMS counts. 
The red stars show more gradual variations with radius, 
and there is little or no correlation with features in the IMS distribution. 
Departures from large scale trends in the RHeB$+$AGB distribution are substantially 
smaller than those in the IMS distribution, amounting to $\sim 0.2$ dex.

	In addition to the periodic nature of structure in the radial distribution 
of IMS stars, there is also a systematic gradient in the relative numbers 
of RHeB$+$AGB and IMS stars. This is investigated in the lower panel 
of Figure 20, where the ratio of objects in the two 
samples is shown. There is a tendency for the 
numbers of RHeB$+$AGB and IMS sources to become more equal towards 
larger R$_{GC}$, indicating that star formation was more centrally 
concentrated in M31 during intermediate epochs than at present. 
This trend is most noticeable when considering sources along the NE axis; while 
a gradient is seen out to R$_{GC} \sim 10$ kpc along the SW axis, it may reverse 
at larger radii. We note that the FUV--NUV color in M31 decreases (i.e. 
becomes bluer) towards larger radii in the Inner Disk when R$_{GC} > 2$ kpc
(Thilker et al. 2005), and this is broadly consistent with the relative radial 
distributions of the RHeB$+$AGB and IMS objects in our data. The change in UV 
color in the interval R$_{GC} \leq 10$ kpc corresponds to a change in 
luminosity-weighted SSP age from $> 450$ Myr to 290 Myr (Thilker et al. 2005).

	Ferguson et al. (2002) investigate the distribution of RGB and AGB stars 
in the outer regions of the M31 disk. Their study intentionally 
avoids the main body of the disk where crowding is an issue for 
their aperture photometry measurements, and so
there is only modest spatial overlap between our MegaCam and the 
Ferguson et al. (2002) datasets. Still, we note that Figure 2 of Ferguson et 
al. (2002) shows filamentary structure at large radii. While 
contamination from background galaxies accounts for a progressively larger 
fraction of the objects in Figure 19 as one moves to larger R$_{GC}$, 
there are hints of structures at large R$_{GC}$ in Figure 19. These will 
prove to be promising targets for future deep photometric studies.

\section{DISCUSSION \& SUMMARY}

	Images obtained with the CFHT MegaCam of the 
Local Group galaxy M31 have been used to examine the spatial distribution and 
photometric properties of the brightest stars in this system. M31 
is an important target for studies of galaxy evolution not only 
because it is the closest external large disk galaxy, but also 
because it has anemic star-forming activity at the 
present day, and is the nearest example of a red disk that is evolving in 
the so-called `green valley'. With four MegaCam pointings we cover most of the M31 
disk out to 100 arcmin (R$_{GC} \sim 20$ kpc) along the major axis, 
allowing us to investigate large-scale trends in stellar content.
The primary goals of our study are (1) to probe the SFH of M31 during the 
past few hundred Myr, and (2) to search for evidence of recent interactions.

	When considered alongside studies of the outer 
regions and bulge of M31, which have yielded insights into 
the evolution of the galaxy during early and intermediate epochs, the 
results of the current work help to develop an understanding of the 
evolution of M31 in the context of other nearby galaxies. M31 has been 
subjected to interactions with companions (e.g. \S 1), and this 
activity has continued to cosmologically recent epochs. 
These interactions have influenced the large scale re-distribution 
of stars and gas, and have had a major influence on the SFH of the galaxy. 
In the remainder of this section we discuss our four principal results:
(1) the nature of the SFH during the past 100 Myr, focusing on the apparent 
rise in the SFR over the last $\sim 10$ Myr that has occured throughout much of 
the galaxy, (2) the spatial distribution of stars that formed 10 -- 100 Myr in 
the past, (3) the distribution of evolved red stars that formed 
$100+$ Myr ago, and (4) the subdued level of star-forming activity at the 
present day, and the evolutionary status of M31.

\subsection{The Recent SFH: Evidence for a Rise in the SFR}

	Stellar density changes with location in M31, and 
this affects the spatial coverage of the stellar 
samples used here to probe the SFH. Because they are among the brightest 
stars in the galaxy, and have relatively blue colors when compared with the 
underlying disk, main sequence stars with ages $\sim 10$ Myr are 
detected over a large fraction of the M31 disk. While resolving fainter 
stars is more challenging, main sequence stars with ages $\sim 100$ Myr 
are still detected at R$_{GC} \geq$ a few kpc in our data. The recent (i.e. 
within a hundred Myr) SFH can thus be investigated over much of the M31 disk. 

	The $u'$ LF of main sequence stars 
are indicative of a constant SFR during the past 10 -- 100 Myr, 
and so suggest that any major disturbance to the disk of M31 during intermediate 
epochs probably occured more than $\sim 100 - 200$ Myr in the past. 
However, the number counts of the brightest main sequence stars indicate that 
the SFR has increased by a factor of $2 - 3$ during the past $\sim 10$ Myr. 
That this upturn in the SFR was not restricted to Ring 10 and Ring 14, 
which are the areas of M31 that contain a large fraction of its young stars 
and star-forming material, indicates that the trigger was an event that 
spurred activity throughout the disk, and is probably not stochastic in nature. 

	A rise in star-forming activity during the past 10 Myr is not seen at 
R$_{GC} > 21$ kpc, and this provides clues into the nature of any event that 
may have triggered the change in the SFR. If the SFR was elevated by an 
interaction with a companion, then the point of closest encounter between M31 
and this companion would have been R$_{GC} < 21$ kpc (e.g. Hopkins et al. 
2009). If the companion passed through the disk plane then the 
point of impact must then have been in the present-day star-forming disk.
Gordon et al. (2006) models an interaction with M32, in which 
that galaxy passed through the inner disk of M31 roughly 20 Myr ago. The timing of 
that event is not very different from our estimate 
of a recent upswing in the disk SFR.

	Large scale spiral structure is often associated with recent interactions. 
If M32 (or another companion) passed through the disk of M31 in the recent 
past then why is there no large scale spiral structure? The amplitude of any 
spiral pattern depends on the mass of the perturber and 
the cold gas content of the larger galaxy. In 
addition, interactions may spur spiral structure only over limited areas, and even 
then these will survive only over dynamical time scales (Revaz et al. 2009). 
If M31 was depleted of cold gas -- or the cold gas distribution was disrupted --  
prior to the most recent interaction with a companion, then prominent spiral 
structure may not be expected.

	The SFH measured from the MegaCam images is broadly consistent with that 
presented by Williams (2002; 2003). Williams (2002) examined 
a suite of HST/WFPC2 images to investigate the SFHs 
of the M31 disk and its environs. Given the widely distributed 
locations of these fields, coupled with HST/WFPC2's modest science field, 
it is perhaps not surprising that there is a substantial field-to-field dispersion 
in the SFHs found from these data. Still, the SFHs for the majority of his fields 
are consistent with a cSFR during the past few hundred Myr, in agreement with the 
SFH deduced from the MegaCam $u'$ LFs of main sequence stars during 
the time interval 10 -- 100 Myr.

	Not all of the HST/WFPC2 fields show evidence of an upturn in 
the SFR within the past 10 -- 20 Myr. This is perhaps not surprising, as 
the distribution of young stars, UV light, MIR emission, and molecular material 
all indicate that the recent rise in star-forming activity occured in isolated 
areas that are distributed throughout the disk. The SFHs deduced 
from the MegaCam data are based on stellar content that 
covers substantial swaths of the M31 disk. The sampling of small star-forming 
areas by the substantially smaller HST/WFPC2 field will be less complete.

	Williams (2003) used wide-field ground-based images that cover a total 
area of 1.4 degrees$^2$ to probe the recent SFH of the M31 disk. The level of 
star-forming activity was found to decline by 50\% from $\sim 250$ Myr to 
$\sim 50$ Myr in the past, although the SFR during the 
past 64 Myr is steady to within $\pm 0.1$ M$_{\odot}$ year$^{-1}$, 
thereby approximating a cSFR to within $\pm 20\%$. The SFR was also found to 
increase during the past 25 Myr, in broad agreement with the results found 
here. The qualitative agreement with the Williams (2003) 
results are significant, as Williams examined a much smaller portion of the 
disk, and used a CMD-fitting technique that relies on stars spanning a range of 
evolutionary states, in contrast to our study which is based on main sequence stars.

\subsection{Signatures of Recent Interactions: Characterizing Coherent Structures in the Disk of M31}

	Many bright main sequence stars in M31 are found in ring- and shell-like 
structures. These rings have a radial periodicity of $\sim 5$ kpc and -- based on 
main sequence stars that formed within the past 100 Myr -- a stellar density that 
is an order of magnitude higher than in the inter-ring regions (Figure 20). 
Such a distribution of young stars indicates that much of the cool gas in M31 has 
been displaced systematically throughout M31, as is evident in the 
distribution of HI and molecular material (e.g. Emerson 1974; Nieten et al. 2006).

	The distribution of objects in the IMS sample indicates that 
the rings have been in place (and forming stars) for at least 
$\sim 100$ Myr. GMCs in the disks of `normal' star-forming galaxies play a key 
role in kinematically heating stars over time spans of a few tens of Myr. The 
modest physical width of the rings in M31 ($\pm 1$ kpc; Figure 20) 
suggests that the orbital angular momenta of GMCs in these 
rings must have a compact distribution, reducing the incidence of collisions and 
thus contributing to the longevity of these structures.

	The IMS sample contains main-sequence stars that formed
over the entirety of the past $\sim 100$ Myr, and this might skew age 
estimates. Still, the majority of objects in the IMS 
sample likely have ages that are not much younger than 100 Myr. Indeed, 
there has been on-going star formation throughout the disk of M31 during the past 
100 Myr (\S 5), which proceeded at a pace that is consistent with a cSFR for 
$t > 10$ Myr. Given this SFH and the relation between age and MSTO magnitude, 
then the majority of objects in the IMS sample will likely have ages that are 
weighted towards the older, as opposed to younger, end of the age range covered 
by the part of the $(M_{g'}, u'-g')$ CMD that contains IMS stars.

	Interactions can produce the ring-like distribution of young stars in 
the M31 disk. Hammer et al. (2010) find that a gas ring forms in models in 
which M31 merges with a large companion on a polar orbit. 
Block et al. (2006) discuss a model in which a companion passed 
through the central regions of M31 $\sim 200$ Myr ago. A propagating density wave, 
similar to that in the Cartwheel galaxy, then triggers star formation as it works 
its way outwards. Block et al (2006) suggest that the companion that spurred 
this event was M32, accompanied by a massive halo to enhance the
strength of the interaction. However, the orbit that they assign
to M32, together with its velocity, places it behind M31, whereas
there is strong evidence that M32 lies in front (Ford et al. 1978). In any event, 
the spatial pattern of the IMS sample argues against this model.
Rather than commencing at its current location only within the last 10 Myr or so,
we find that the star formation in Ring 10 has existed for at least 
100 Myr. The existence of similar long-lived star-forming structures 
outside Ring 10, such as Ring 14, is further evidence against a 
propagating wave origin for Ring 10.

	In order to have a cSFR during the past $\sim 100$ Myr, it is likely 
that the last major interaction between M31 and a companion occured at least 
$\sim 500$ Myr in the past. This leads us to 
speculate that the displacement of cool gas into rings may be 
an artifact of the event that triggered the last major 
upswing in star-forming activity 1 -- 2 Gyr ago. The event that produced this 
rise in star-forming activity undoubtedly affected the ISM of M31 given the 
large population of globular clusters that formed at this time (Puzia et al. 2005). 

	Hammer et al. (2010) argue that the 10 kpc ring may have resulted from a 
merger between M31 and a large companion more than 5 Gyr ago. 
There are other possible perturbers: M33 is one, while another is M32 (or its 
progenitor). M32 is of interest as it contains a substantial population of 
stars with an age of a few billion years (e.g. Davidge 1990b; Davidge 2000; del 
Burgo et al. 2001; Worthey 2004; Rose et al. 2005) that are uniformly 
distributed throughout the galaxy (Davidge \& Jensen 2007). The 
properties of M32 are also consistent with it having been sculpted 
by interactions. Compact elliptical galaxies like M32 are 
associated with dense environments, and there is evidence that the morphology of 
M32 may have changed dramatically with time. Indeed, remnants of a 
fossil disk have been detected around M32 (Graham 2002; Choi et al. 2002), 
suggesting that the present-day galaxy started out as a much larger disk 
system. M32 also lacks an entourage of globular clusters, as well as interstellar 
gas (Sage et al. 1998) and dust (Gordon et al. 2006). These components may have 
been removed if M32 was subjected to tidal stripping. The 
total mass of M32 is only $\sim 2\%$ that of M31, raising the question if it 
would be large enough to influence star formation in M31, although if it 
interacted with M31 then its initial mass was probably larger than at the present 
day. Simulations suggest that satellites with masses that are 10\% of the larger 
system can significantly affect disk properties (Mori \& Rich 2008; 
Qu et al. 2011). 

	The events that re-distributed gas and spurred elevated levels of star 
formation may leave signatures near the disk boundary. Mergers are expected to 
move the boundaries of stellar disks outwards (e.g. Younger et al. 2007; Cook 
et al. 2009), as gas that loses angular momentum and moves inwards scatters stars 
to larger radii, pushing out the radial extent of the disk. In fact, an extended 
disk is observed in M31 (e.g. Ferguson \& Johnson 2001), and is traced 
out to $\sim 40$ kpc (Ibata et al. 2005). The outer regions 
of the disk account for 10\% of the total M31 disk luminosity (Ibata 
et al. 2005), indicating that it is a significant structural component. 
Ferguson \& Johnson (2001) also conclude that the 
majority of stars in the outer disk have ages in excess of a Gyr, which is 
consistent with a population of stars that were displaced outward during the 
event that produced the population of intermediate age globular clusters. 
In contrast, Ibata et al. (2005) and Pe\~narrubia et al. (2006) 
argue that this part of the disk is the product of satellite 
accretion, rather than the re-configuring of the orbits of stars that originated 
at smaller radii.

\subsection{Signatures of Recent Interactions: The Asymmetric Distribution of Evolved Red Stars}

	Lop-sided structure in galaxies is not rare (e.g. Zaritsky \& Rix 1997), 
and may be a signature of the large-scale radial movement of gas in disks (e.g. 
Reichard et al. 2009), or the non-uniform accretion of cosmic gas (Bournaud et 
al. 2005). The projected density of RHeB$+$AGB stars with R$_{GC} < 10$ kpc 
along the NE segment of the major axis is roughly two times that along the 
SW segment. Given that these stars have ages 100 -- 200 Myr, then the 
lop-sided distribution was likely spurred by an event that occured within 
the past $\sim$ 0.5 Gyr. Such a time scale would allow an elevated 
SFR to die down so that a cSFR could be seen in the time interval probed 
by the $u'$ LF of main sequence stars in the Inner Disk.

	The asymmetric distribution of RHeB$+$AGB stars is consistent with other 
observations. Williams (2003) found that the region $\sim 20 - 30$ arcmin of the 
north east of the galaxy center was an area of enhanced star-forming activity 
64 -- 256 Myr in the past. In addition, circumstellar envelopes around massive AGB 
stars contribute significantly to the MIR light from systems 
containing stars that formed within the past few Gyr. 
The [5.8] and [8.0] major axis profiles of M31 shown in Figure 2 
of Barmby et al. (2006) have higher levels of emission between $\sim 800$ and 1800 
arcsec (13 -- 30 arcmin) along the NE major axis segment than the corresponding 
section of the SW segment. Finally, Davidge (2012) examined 2MASS images and 
discovered a substantial concentration of AGB stars $\sim 3.5$ kpc NE of the 
center of M31. This structure is populated by stars with ages in excess of 100 Myr.

	The presence of a large-scale asymmetry in the distribution of stars with 
ages 100 - 200 Myr is perhaps surprising given that 
rotational shear might be expected to blur such structures. 
However, shear will not be an issue for stars at radii where the disk 
kinematics follow solid body rotation. While rotation curves derived 
from HI suggest that significant shear might be expected even in the inner 
regions of M31 (e.g. Braun 1991), the gas dynamics at small radii may not reflect 
those of objects in the stellar disk (e.g. Kent 1989a). 
Still, there is agreement between the rotational properties of M31 defined by 
HI and PNe at R$_{GC} \sim 12$ kpc (Halliday et al. 2006), and the 
rotation velocity measured from HII regions drops near 25 -- 30 armin (R$_{GC} 
\sim 5 - 6$ kpc; Kent 1989b). For comparison, the asymmetry in the RHeB$+$AGB 
number counts disappears when R$_{GC} > 10$ kpc. Kinematic measurements of bright 
AGB stars in the inner disk of M31 will help establish if they are in 
the disk plane. Such measurements could be made in the 
$K-$band, where the brightest AGB stars in M31 have $K \leq 17$ and significant 
gains in angular resolution can be realised with adaptive optics systems.

	The event that produced the asymmetric 
distribution of RHeB$+$AGB stars may have left signatures 
in the extraplanar regions if it involved an encounter with a satellite. 
The GSS, which Mori \& Rich (2008) model as the result of a moderate mass companion 
passing through the M31 disk, is one candidate signature of such an event. 
Fardal et al. (2007) find a kinematic age for the GSS of 750 Myr, 
which is a few hundred Myr older than the RHeB$+$AGB stars examined here.

	The dwarf elliptical galaxy NGC 205 is one of the largest companions of 
M31, and has a projected distance of only $\sim 10$ kpc from the 
center of M31. Twisting of the outer isophotes of NGC 205 is 
consistent with tidal disruption (Choi et al. 2002). 
There is evidence in the resolved stellar content of NGC 205 for a large
episode of star formation $\sim 100$ Myr in the past, and another 
a few hundred Myr before that (Davidge 2003). 
The timing of the first event coincides with the last M31 disk-crossing predicted 
by Cepa \& Beckman (1988), while the second is consistent with their period of NGC 
205 about M31. The second star-forming episode is also consistent with the
timing of the event that may have produced the lop-sided distribution 
of RHeB$+$AGB stars in M31. 

	Howley et al. (2008) have re-examined the orbital properties 
of NGC 205 and suggest that it might be on its first passage by M31, counter to the 
orbit proposed by Cepa \& Beckman (1988). If NGC 205 has interacted with 
M31 then a tidal tail would be expected, and McConnachie et al. (2005) report the 
possible detection of such a structure. Given that there is no consensus 
or firm evidence on the orbit of NGC 205 about M31 then the connection 
between this satellite and the larger galaxy must remain speculative for now.

\subsection{M31 as a Red Disk Galaxy}

	Photometric surveys of galaxies reveal distinct 
red and blue sequences on CMDs (e.g. Strateva et al. 2001; Bell et al. 2004). 
The majority of objects in the blue sequence are disk-dominated, while the 
majority of objects in the red sequence are spheroid-dominated (Strateva et 
al. 2001). Mutch et al. (2011) examine the morphology and integrated colors of M31, 
and conclude that it is currently in the `green valley' that falls between the blue 
and red sequences. They also conclude that at least one in six 
nearby disk galaxies, including the Galaxy, lie in the green valley or on the 
red sequence. If the SFR of M31 has recently increased by a 
factor of $2 - 3$, then the appearance of M31 $\sim 100$ Myr in the past M31 would 
have been different from the present day. If the current SFR is one 
third that of M33, for which the SFR is 0.5 -- 0.7 M$_{\odot}$ 
year$^{-1}$ (Hippelein et al. 2003; Verley et al. 2009), then the SFR in M31 
would have been only 0.1 M$_{\odot}$ year$^{-1}$
a hundred Myr ago. When considered in the context of star-forming activity, M31 
at that time would have appeared as a largely dead system.

	The sSFR in M31 is similar to that in nearby red disk galaxies, 
such as the lenticular galaxy NGC 5102. Using the integrated $K-$band brightnesses 
of M31 and NGC 5102 from Jarrett et al. (2003) as proxies for computing 
relative stellar masses, and assuming a SFR of $\sim 0.02$ M$_{\odot}$ 
year$^{-1}$ for NGC 5102 (Davidge 2008), then the sSFR of M31 
100 Myr ago would have been comparable to that of NGC 5102 at the present day. 
It is perhaps worth noting that M31 and NGC 5102 share common morphological 
properties. Young stars in NGC 5102 have a ring-like distribution and 
the lop-sided distribution of AGB stars in NGC 5102 (Figure 5 of Davidge 
2010) is reminiscent of the distribution of RHeB$+$AGB stars in M31. 

	Bundy et al. (2010) examine the morphologies 
of red sequence galaxies, and find that a significant number are passively evolving 
disks where large-scale star formation has ceased.
The red disk galaxies examined by Bundy et al. (2010) 
tend to have more prominent bulges than blue, star-forming disk galaxies. 
Red disk galaxies are found in a range of environments, suggesting that 
they are not the exclusive product of evolution in high density regions.
The stellar archeology of M31 may provide clues into how red 
disk galaxies evolve in environments like the Local Group. 

	Star formation in galaxies is influenced by a number of factors, one of 
which is environment, and galaxies with companions tend to have high 
sSFRs (Li et al. 2008). The sSFR also depends on galaxy mass, in the sense that 
the sSFR at the present day rises as one moves to 
lower masses (e.g. Kauffmann et al. 2003). Bundy et al. (2010) suggest that star 
formation in red disks may have been halted by mergers and subsequent interactions. 
These events would have driven the rapid consumption of star-forming material, and 
ultimately contributed to the disruption of the cold ISM. 
Red sequence disks are expected to evolve into pressure-supported systems if they 
experience a major merger, as there is no cool gas from which a disk could re-form. 

	To the extent that M31 can be considered to be a typical red 
disk galaxy, its past history appears to favor the `two stage' model 
forwarded by Bundy et al. (2010), in which a disk is first stripped of gas, and 
is then pummeled by other systems to produce a red galaxy that is 
supported mainly by random motions, rather than rotation. The final 
pressure-supported configuration of M31 may be realised a few Gyr in the future, 
when it is projected to merge with the Galaxy (e.g. Cox \& Abraham 2008). 
Barring an injection of cool gas into the M31 disk during the next few Gyr, 
this merger will occur when M31 is gas poor and -- based on the present-day modest 
SFR -- contains a preponderance of stars with ages in excess of a few Gyr.

\acknowledgements{GFL thanks the Australian Research Council for support through his
Future Fellowship (FT100100268) and Discovery project (DP110100678). Financial 
support for this work was provided in part by the POMMME project 
(ANR 09-BLAN-0228). We thank the anonymous referee for comments that improved the 
manuscript.}

\parindent=0.0cm

\clearpage

\begin{table*}
\begin{center}
\begin{tabular}{lcc}
\tableline\tableline
Field \# & Date Observed & Exposure times \\
 & & (seconds) \\
\tableline
1 & Sept 13, 2010 & $9 \times 600 (u*)$ \\
 & & $4 \times 200 (g')$ \\
 & & $4 \times 200 (r')$ \\
 & & \\
2 & Oct 31, 2010 & $6 \times 600 (u*)$ \\
 & & $4 \times 200 (g')$ \\
 & & $4 \times 200 (r')$ \\
 & & \\
3 & Nov 1, 2010 & $6 \times 600 (u*)$ \\
 & & $4 \times 200 (g')$ \\
 & & $4 \times 200 (r')$ \\
 & & \\
4 & Oct 30, 2010 & $9 \times 600 (u*)$ \\
 & & $4 \times 200 (g')$ \\
 & & $4 \times 200 (r')$ \\
 & & \\
\tableline
\end{tabular}
\end{center}
\caption{Log of Observations}
\end{table*}

\clearpage

\begin{table*}
\begin{center}
\begin{tabular}{rccc}
\tableline\tableline
R$_{GC}$ & $\overline{u'-g'}$ & $\sigma_{u'-g'}$ & $\tau_{g'}$ \\
\tableline
6 kpc & $0.380 \pm 0.003$ & $\pm 0.151$ & 0.6 \\
8 kpc & $0.384 \pm 0.002$ & $\pm 0.151$ & 0.6 \\
10 kpc & $0.378 \pm 0.001$ & $\pm 0.154$ & 1.0 \\
12 kpc & $0.377 \pm 0.002$ & $\pm 0.140$ & 0.8 \\
14 kpc & $0.383 \pm 0.002$ & $\pm 0.144$ & 0.4 \\
\tableline
\end{tabular}
\end{center}
\caption{Photometric Properties of the Blue Plume Near $g' = 22$}
\end{table*}

\clearpage

\begin{figure}
\figurenum{1}
\epsscale{1.25}
\plotone{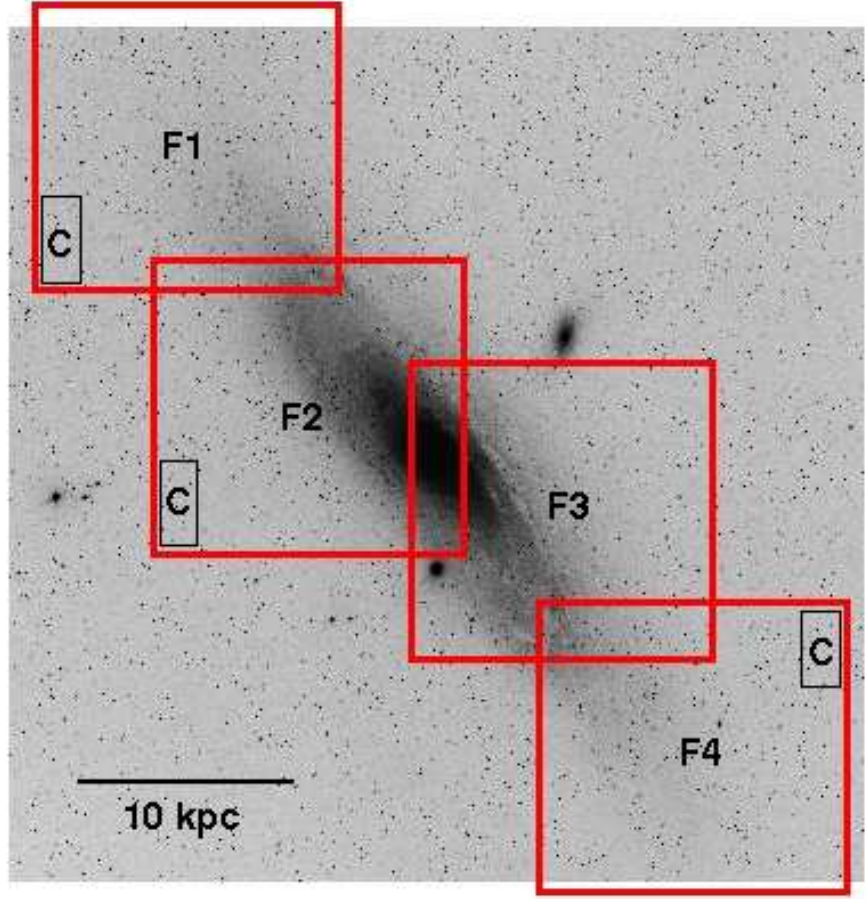}
\caption{The locations of the four MegaCam fields. The 
reference image is from the DSS, and is in the E band. North is at the top, 
and East is to the left. Each MegaCam field covers roughly $\sim 1$ degree$^2$. 
The locations of the three control fields that are used to estimate 
contamination from foreground and background objects are also indicated.}
\end{figure}

\clearpage

\begin{figure}
\figurenum{2}
\epsscale{0.75}
\plotone{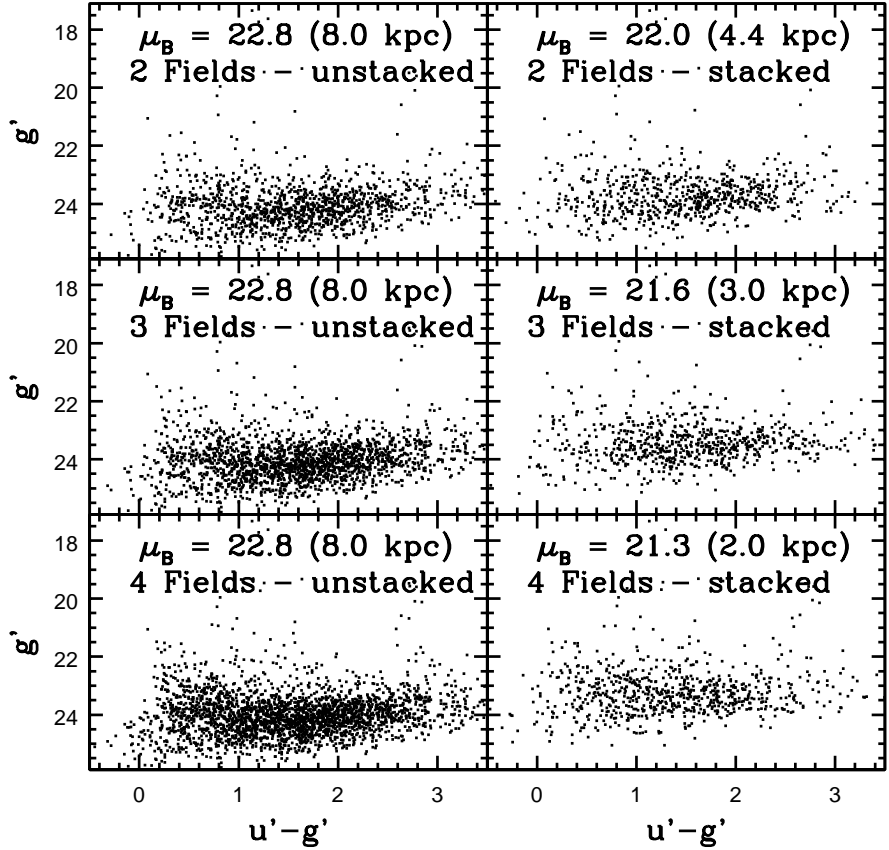}
\caption{Results of image stacking experiments for $u' + g'$. The left hand column 
shows the $(g', u'-g')$ CMDs of two (top row), three (middle row), 
and four (bottom row) $500 \times 500$ pixel$^2$ regions in Field 1. The right 
hand column shows the CMDs that are obtained after images of these same 
regions are stacked to simulate areas of higher stellar 
density. A comparison of the CMDs in the two columns 
allows the impact of crowding on the photometry 
to be assessed. Not surprisingly, the faint limit of the CMDs rises with 
increasing $\mu_B$, while sample incompleteness at a given magnitude near the 
faint limit increases with stellar density. The 
agreement between the unstacked and stacked CMDs when $g' < 23$ 
suggests that the photometric properties of the brightest stars are not affected 
by crowding, even in the inner disk of M31.}
\end{figure}

\clearpage

\begin{figure}
\figurenum{3}
\epsscale{0.75}
\plotone{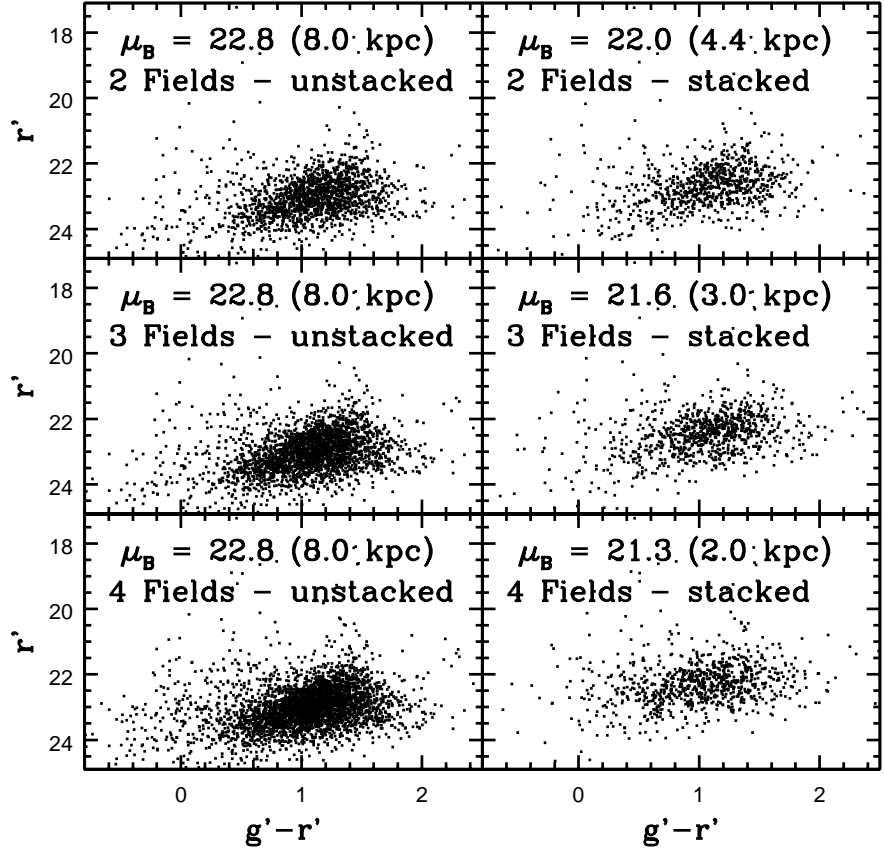}
\caption{The same as Figure 2, but showing the results of stacking 
experiments for $g' + r'$. Crowding affects the CMDs when $r' > 22$.}
\end{figure}

\clearpage

\begin{figure}
\figurenum{4}
\epsscale{1.00}
\plotone{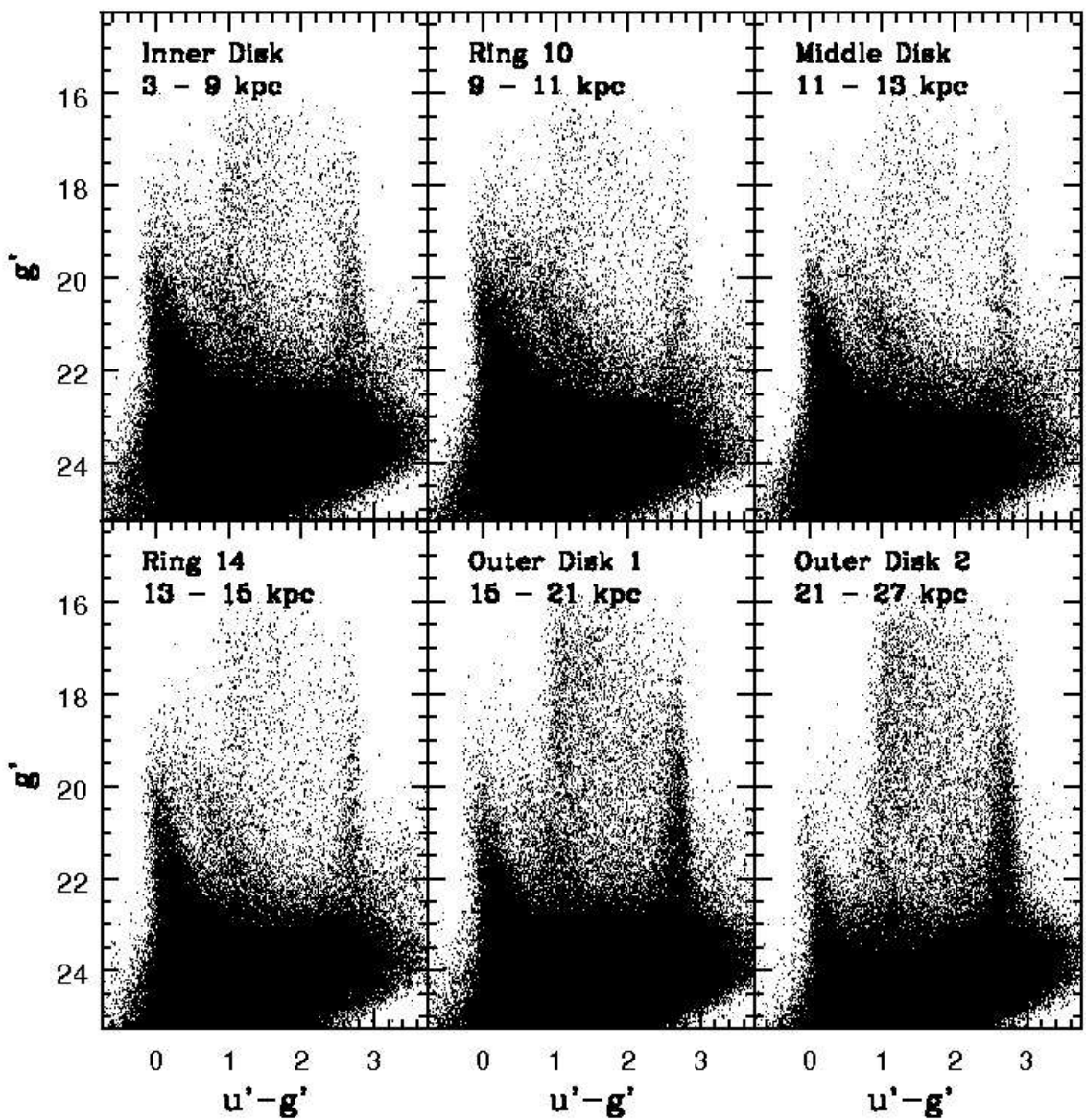}
\caption{The $(g', u'-g')$ CMDs of objects in the disk of M31. The 
distances in each panel are in the plane of the disk, assuming 75$^o$ inclination 
(Gordon et al. 2006). Main sequence stars with $g' < 20$ (M$_{g'} < -5$, which 
have ages $< 10$ Myr) are found out to R$_{GC} \sim 21$ kpc, while fainter 
main sequence stars are traced out to even larger radii.
Galactic foreground stars form the diffuse sequence with $g' > 16$ and $u'-g'$ 
between 1 and 3.}
\end{figure}

\clearpage

\begin{figure}
\figurenum{5}
\epsscale{1.00}
\plotone{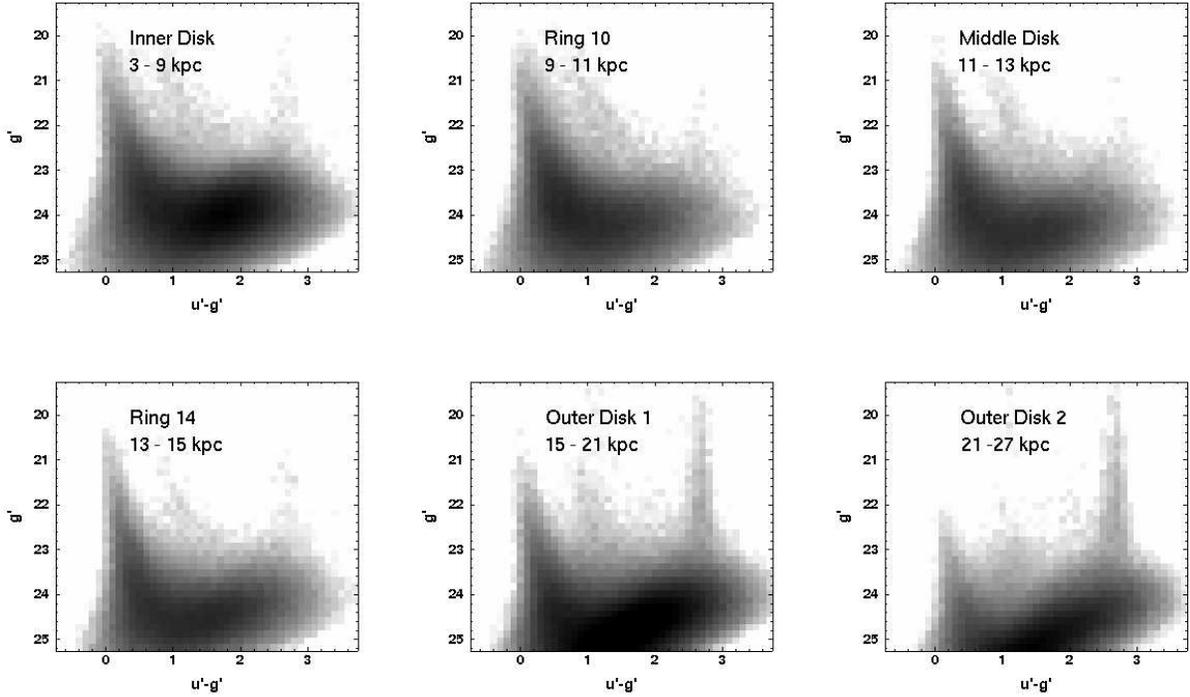}
\caption{Hess diagrams of the $(g', u'-g')$ CMDs. Binning factors of 0.1 
magnitude in $u'-g'$ and 0.15 mag in $g'$ have been employed, and the aspect ratio 
is the same as for the CMDs in Figure 4. There is a prominent collection of objects 
with $u'-g' \geq 1$ and $g' > 23$ in the Inner Disk Hess diagram, which is 
attributed to enhanced star-forming activity in this part of M31 100 -- 200 
Myr in the past (\S 6). The Inner Disk aside, there is a tendency for the faint red 
tongue of objects with $u'-g' > 1$ to become more 
pronounced as R$_{GC}$ increases. This is due -- at least in part -- to increased 
contamination from background galaxies in the diffuse outer regions of the disk.}
\end{figure}

\clearpage

\begin{figure}
\figurenum{6}
\epsscale{0.75}
\plotone{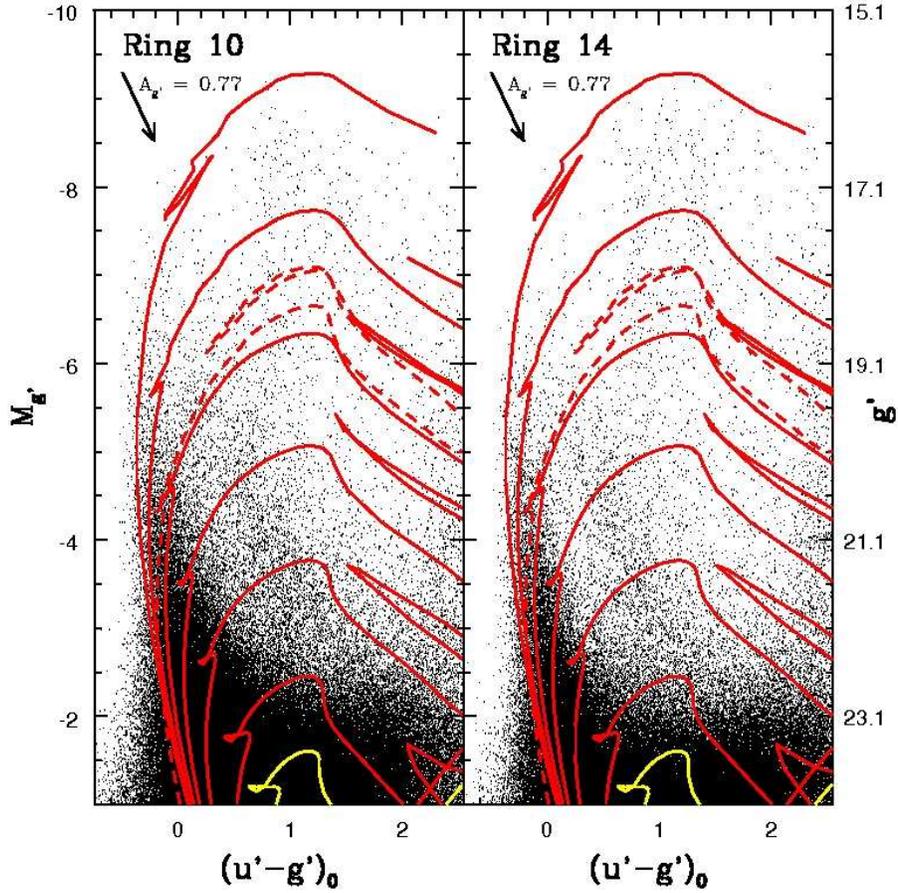}
\caption{The $(M_{g'}, u'-g')$ CMDs of Ring 10 and Ring 14 are compared with 
Z = 0.019 isochrones from Girardi et al. (2004). A reddening vector, with a  
length that corresponds to the reddening correction applied in this study, 
is also shown. The red lines are models with ages 5, 10, 20, 40, 
80, and 160 Myr, while the yellow line is a 240 Myr isochrone. 
The dashed red line is a 20 Myr Z = 0.004 isochrone, which 
is included to demonstrate metallicity effects. The theoretical ZAMS falls along 
the blue envelope of the observed main sequence, while the locus of TAMS 
points tracks the red envelope. Given that the dispersion due to photometric 
errors is modest ($\pm 0.07$ mag at M$_g' = -2$), the agreement with the modelled 
ZAMS and TAMS suggests that the width of the main sequence in our 
data is defined primarily by the ages of stars in our fields, 
rather than by binarity or differential reddening. It is also evident 
that the faint red end of the CMD contains evolved stars with 
ages $\geq 0.2$ Myr that are moving from the main sequence to the AGB.}
\end{figure}

\clearpage

\begin{figure}
\figurenum{7}
\epsscale{1.00}
\plotone{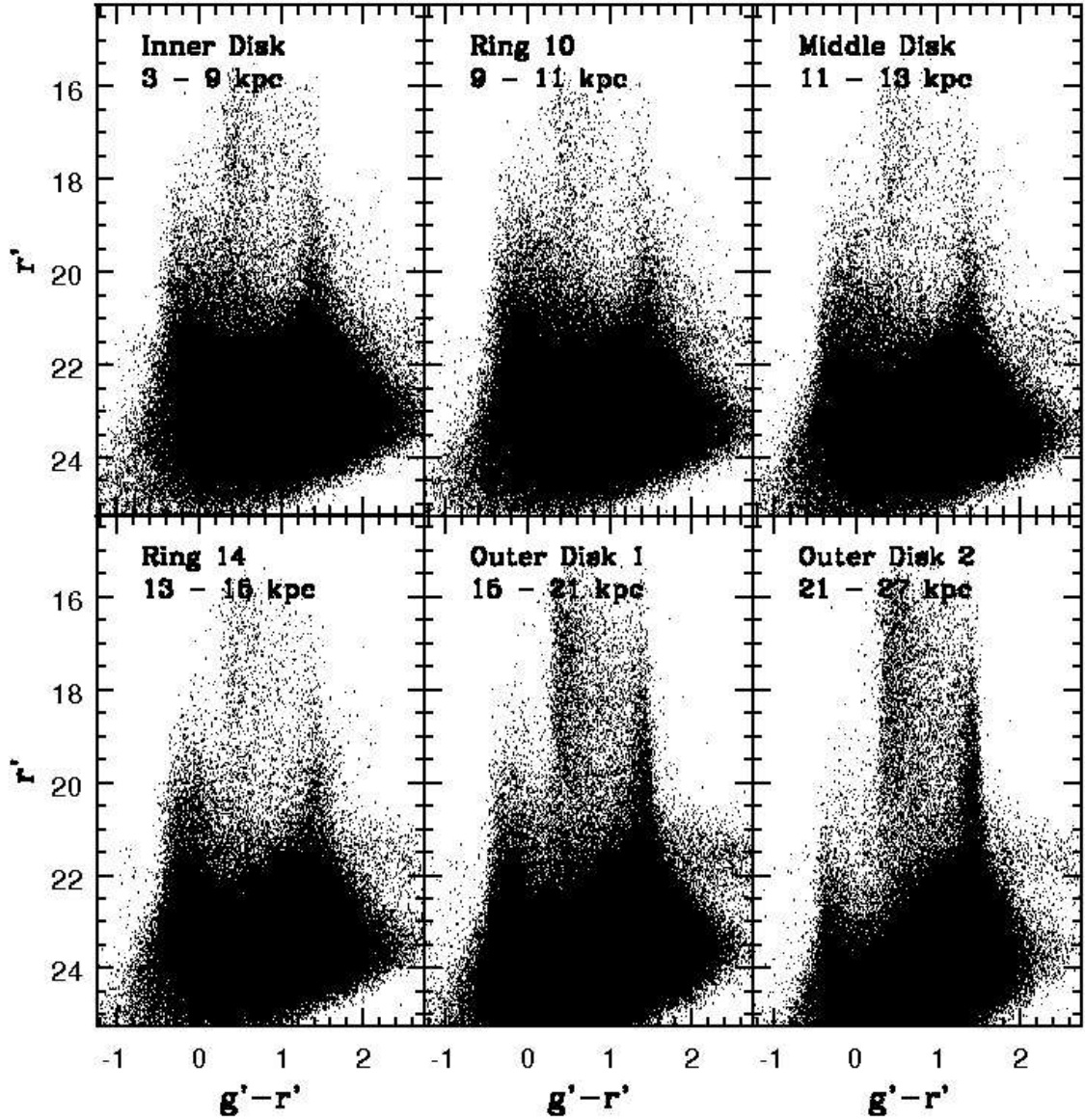}
\caption{The same as Figure 4, but showing $(r', g'-r')$ CMDs. The plume of 
objects with $g'-r' < 0$ is a mix of main sequence stars and BSGs, while 
RHeB stars dominate the collection of objects with $r' > 21$ and $g'-r'$ between 
0 and 2.}
\end{figure}

\clearpage

\begin{figure}
\figurenum{8}
\epsscale{1.00}
\plotone{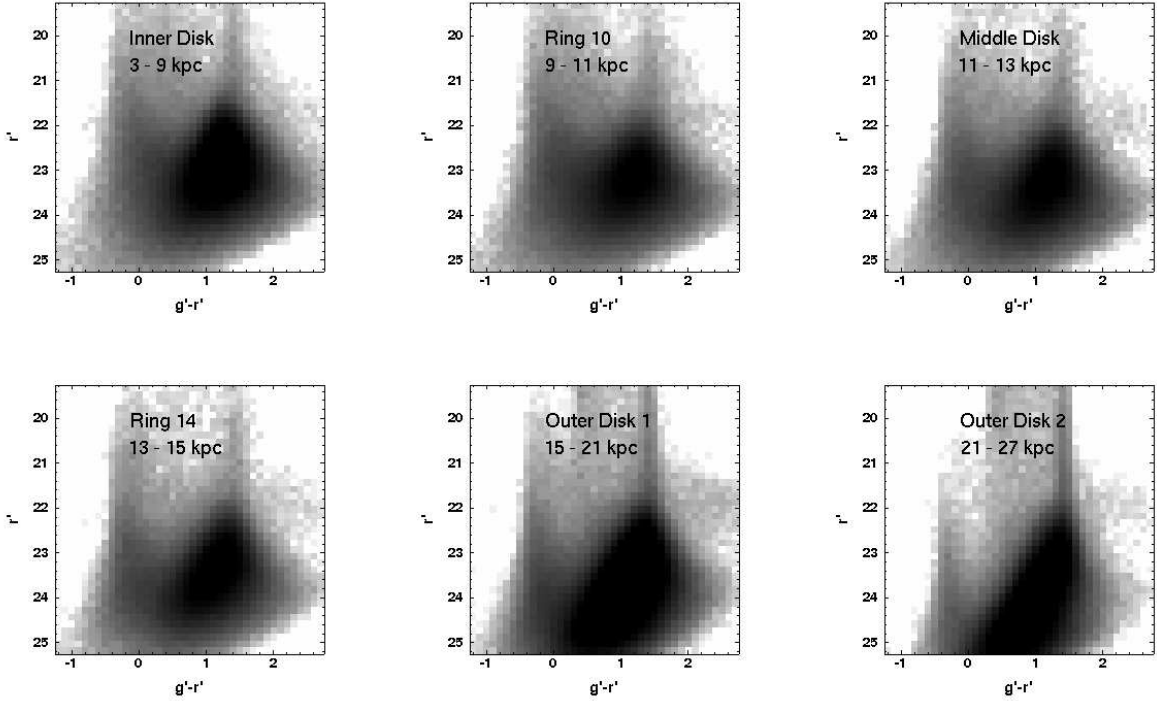}
\caption{Hess diagrams of the $(r', g'-r')$ CMDs. A 0.1 magnitude binning is 
employed in $g'-r'$, while 0.15 mag binning has been employed in $r'$. The 
aspect ratio is the same as the CMDs in Figure 7. There is a prominent red 
sequence with $g'-r' \geq 0.5$ and $r' > 22$ in the Inner Disk, which is made up 
of RHeB and AGB stars and is attributed to enhanced star-forming activity in this 
part of M31 during intermediate epochs (\S 6). Contamination from background 
galaxies becomes more important in this red sequence towards larger R$_{GC}$.}
\end{figure}

\clearpage

\begin{figure}
\figurenum{9}
\epsscale{0.75}
\plotone{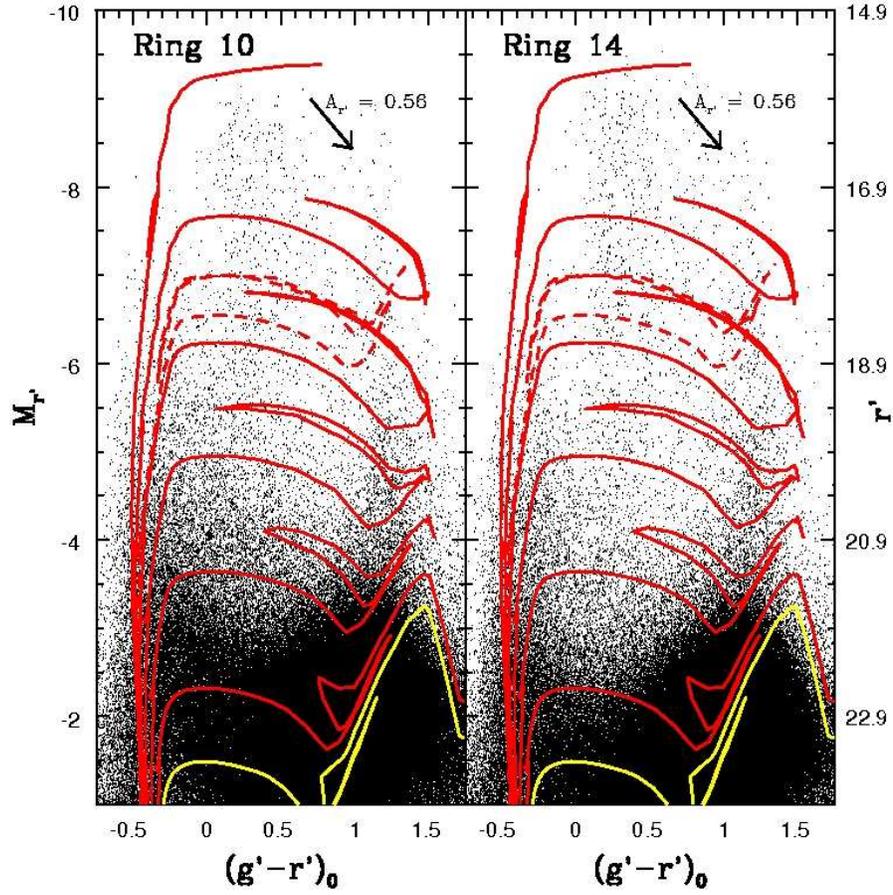}
\caption{The same as Figure 6, but showing $(M_{r'}, g'-r')$ CMDs. 
The near-vertical trajectory of the isochrones after they leave the ZAMS 
produces a more diffuse sequence of blue stars than on the 
$(g', u'-g')$ CMDs. The collection of red objects with M$_{r'} > -3$ is 
populated by intermediate mass RHeB stars, AGB stars, and background galaxies.}
\end{figure}

\clearpage

\begin{figure}
\figurenum{10}
\epsscale{0.75}
\plotone{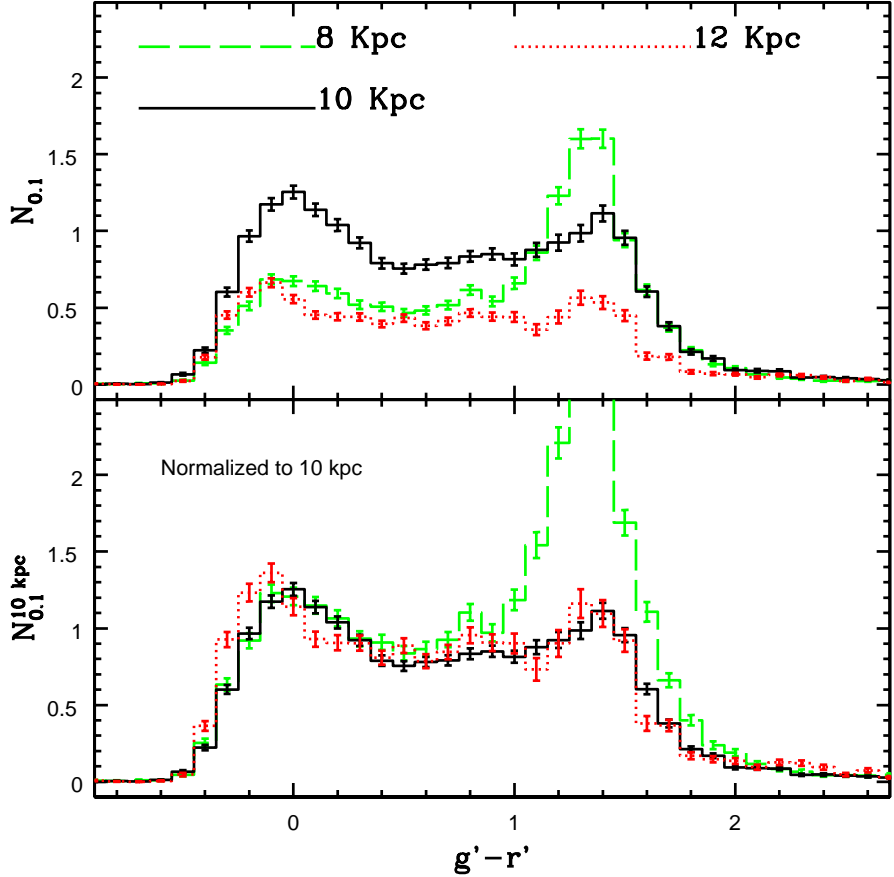}
\caption{The $(g'-r)$ distributions of objects with M$_{r'}$ between --3 and --4 
(i.e. $r'$ between 20.9 and 21.9) in three radial intervals. N$_{0.1}$ 
is the number of objects arcmin$^{-2}$ per 0.1 magnitude interval in $g'-r'$, 
corrected for foreground and background objects using the procedure 
described in the text. The color distributions in the lower panel are the same 
as those in the top panel, and have been normalized to the number of 
objects with $g'-r'$ between --0.25 and 0.25 (i.e. the approximate color interval 
of the blue peak). The peak color of the blue plume in the 12 kpc intervals is 
offset $\sim 0.1$ magnitudes blueward of the peak in the other intervals. Given the 
radial uniformity of the $u'-g'$ color of the main sequence 
(\S 4.1), coupled with the path followed by isochrones on the $(r', g'-r')$ plane, 
we attribute this offset to differences in the SFH 
within the past $\sim 0.1$ Myr. The 8 kpc interval contains 
the highest fractional contribution from evolved red stars, indicating that this 
part of M31 contains a higher fraction of stars that formed $100+$ Myr in the 
past than in the other areas.}
\end{figure}

\clearpage

\begin{figure}
\figurenum{11}
\epsscale{0.75}
\plotone{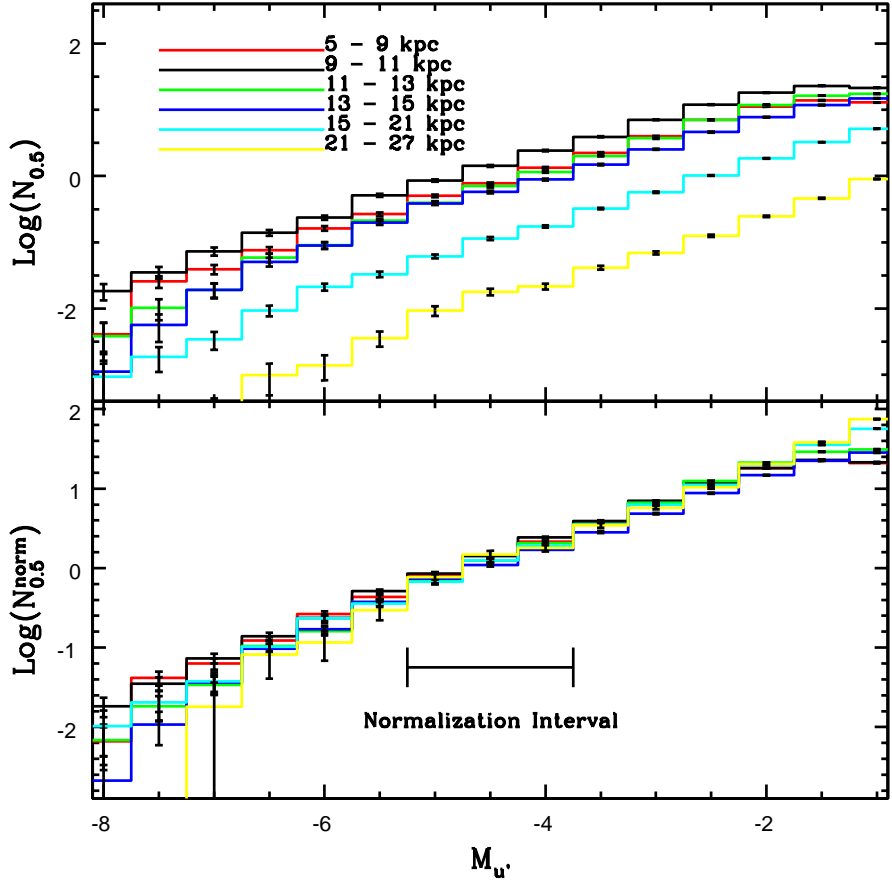}
\caption{(Top panel) The M$_{u'}$ LFs of objects with $(u'-g')_0$ between --0.5 
and 0.5, where N$_{0.5}$ is the number of sources arcmin$^{-2}$ per 0.5 
magnitude interval in $u'$. The LFs have been corrected for foreground and 
background contamination using the procedure described in the text. (Bottom panel) 
The M$_{u'}$ LFs normalized to the Ring 10 counts between M$_{u'} = 
-3.75$ and --5.25. The LFs of sources with R$_{GC}$ between 15 
and 27 kpc differ from those at smaller radii, indicating 
that the mix of massive and intermediate mass stars changes throughout 
the M31 disk.}
\end{figure} 

\clearpage

\begin{figure}
\figurenum{12}
\epsscale{0.75}
\plotone{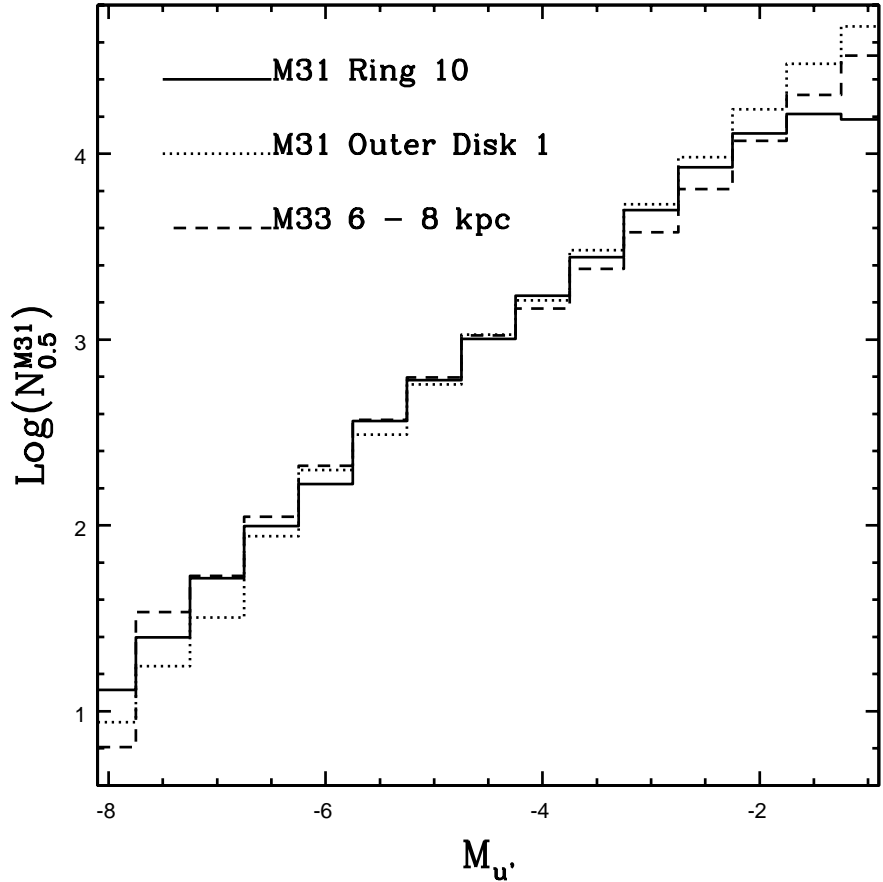}
\caption{The Ring 10 and Outer Disk 1 LFs are compared with the LF of stars 
in the M33 disk with R$_{GC}$ between 6 and 8 kpc. 
The LFs have been normalized to match the M31 Ring 10 LF 
in the M$_{u'}$ interval between --3.75 and --5.25. The M33 LF differs from 
the M31 LFs, indicating different recent SFHs.}
\end{figure}

\clearpage

\begin{figure}
\figurenum{13}
\epsscale{0.75}
\plotone{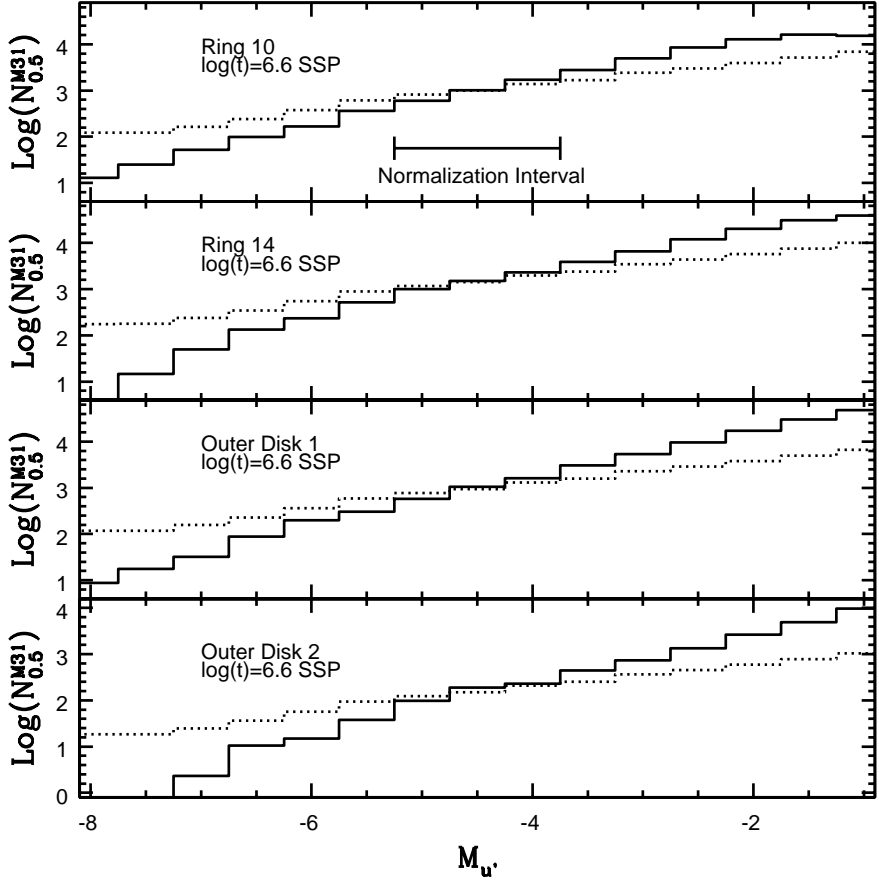}
\caption{The $u'$ LFs of main sequence stars in four radial intervals (solid 
lines) are compared with an SSP model LF having $t = 5$ Myr, Z = 0.019 and an 
IMF exponent $\alpha = -2.7$ (dotted lines). The model has been shifted 
vertically to match the M31 LFs in the interval 
M$_{u'} = -3.75$ to --5.25. N$_{0.5}^{M31}$ is the number of sources in each 
interval arcmin$^{-2}$ per 0.5 magnitude interval in M$_{u'}$. 
The SSP model is a poor match to the observations.}
\end{figure}

\clearpage

\begin{figure}
\figurenum{14}
\epsscale{0.75}
\plotone{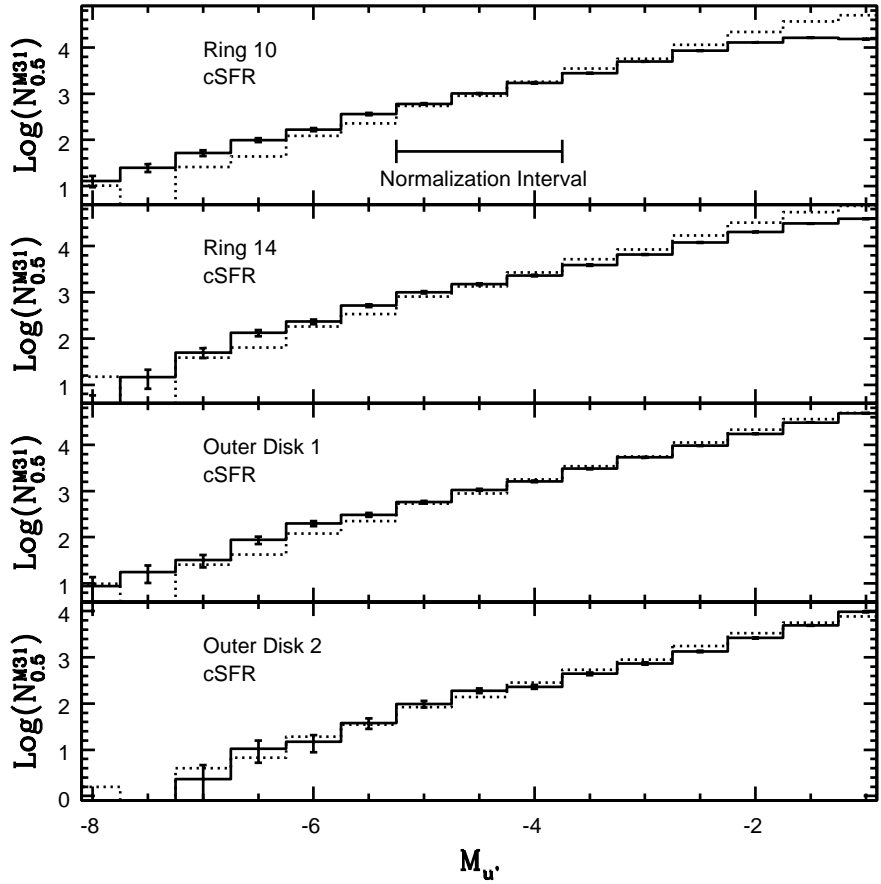}
\caption{The same as Figure 13, but showing constant SFR (cSFR) models (dotted 
lines). The cSFR model is a better match to the observations than the SSP model in 
Figure 13. Still, with the exception of Outer Disk 2, there is a tendency for the 
models to be steeper than the observed LFs.}
\end{figure}

\clearpage

\begin{figure}
\figurenum{15}
\epsscale{0.75}
\plotone{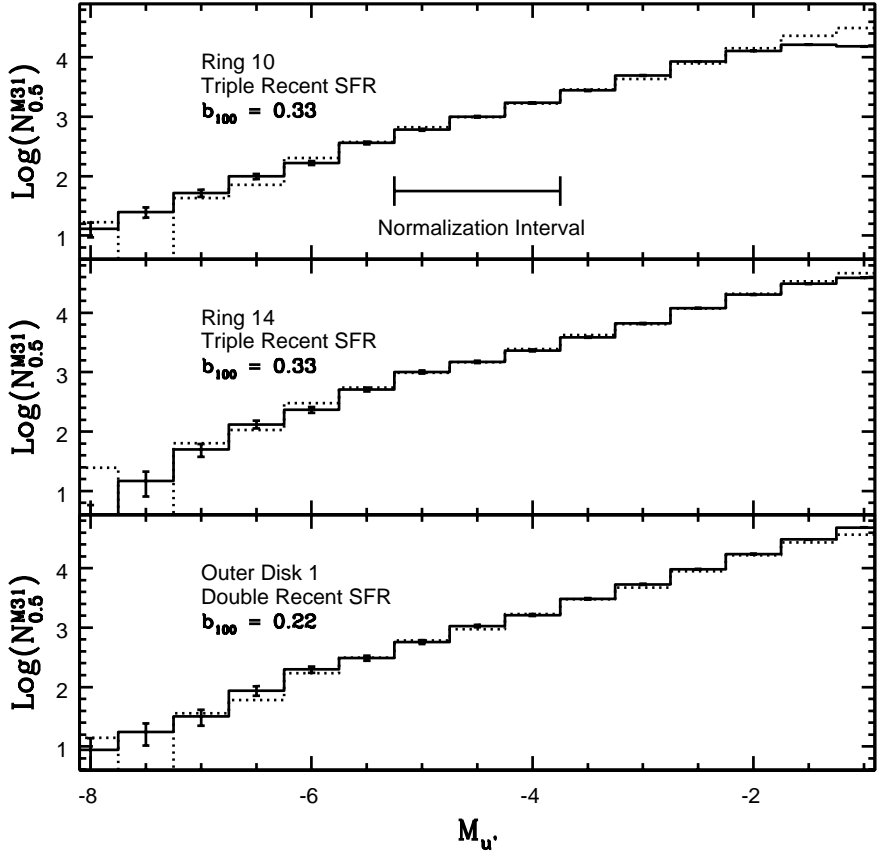}
\caption{The same as Figure 13, but showing models that assume a cSFR for $t > 
10$ Myr, and an increase in the SFR when $t < 10$ Myr. b$_{100}$ is the ratio of 
the number of stars that formed within the past 10 Myr to those that formed 
10 -- 100 Myr in the past; b$_{100} = 0.11$ for a cSFR. The Outer Disk 2 LF is 
not shown, as it is adequately represented by the cSFR model (Figure 14). Models 
with elevated recent SFRs provide a better match to the LFs 
shown here than the cSFR models.}
\end{figure}

\clearpage

\begin{figure}
\figurenum{16}
\epsscale{0.75}
\plotone{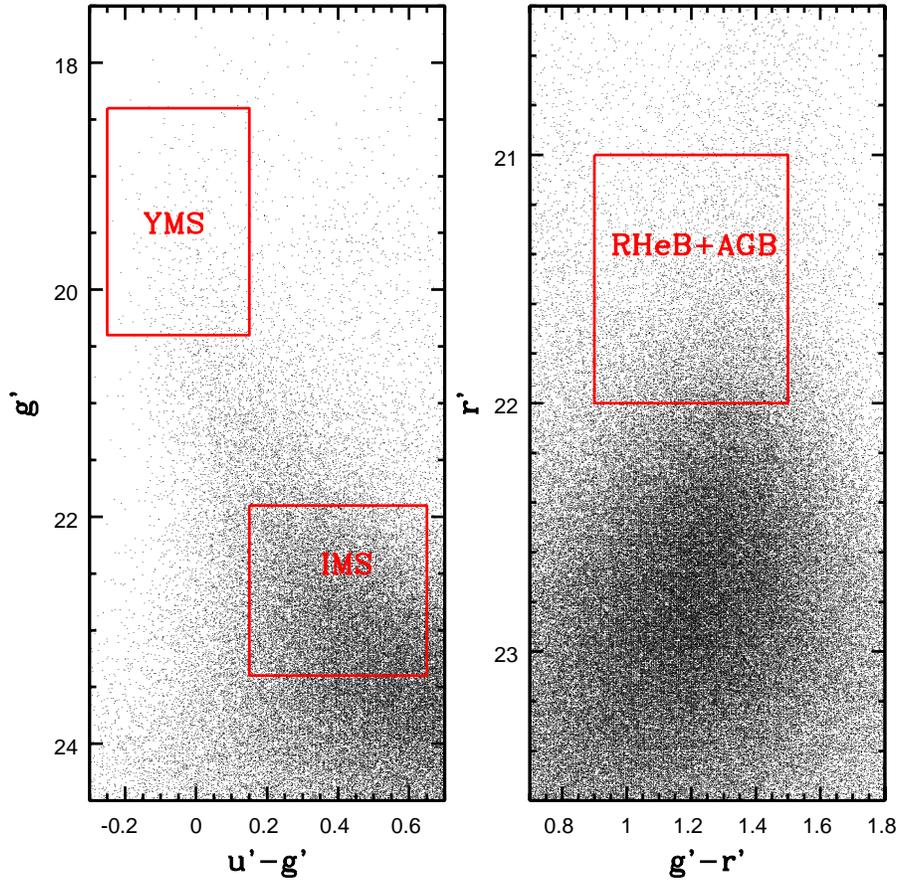}
\caption{Sections of the $(g', u'-g')$ and $(r', g'-r')$ CMDs of Ring 10. 
The areas that are used to identify the samples of young main sequence (YMS), 
intermediate-age main sequence (IMS), and RHeB$+$AGB objects that are examined in 
Figures 17, 18, 19, and 20 are indicated.}
\end{figure}

\clearpage

\begin{figure}
\figurenum{17}
\epsscale{1.00}
\plotone{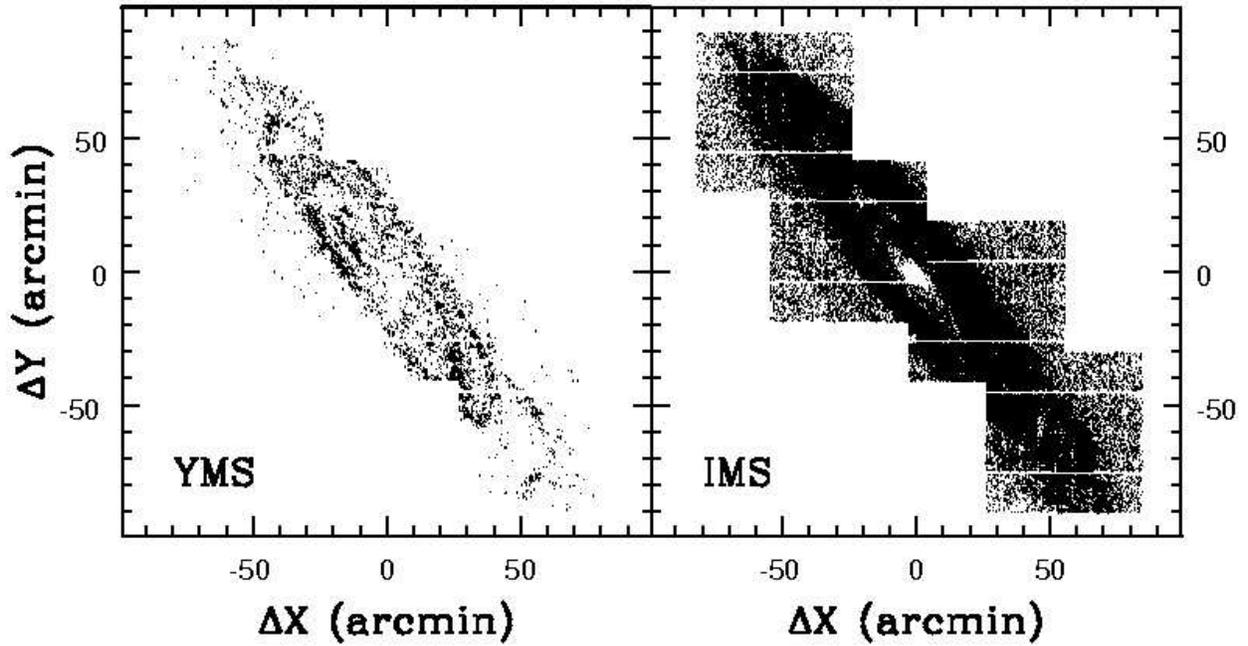}
\caption{The distributions of sources in the 
YMS ($t \geq 10$ Myr) and IMS ($t \sim 100$ Myr) samples as observed on the sky. 
The x and y axes show distances in arcmin from the center of the galaxy. 
Stars in the YMS sample are located throughout the disk, although they 
tend to congregate in rings and arcs. Rings are also evident in the IMS sample, 
indicating that these structures have ages $\geq 100$ Myr.}
\end{figure}

\clearpage

\begin{figure}
\figurenum{18}
\epsscale{1.00}
\plotone{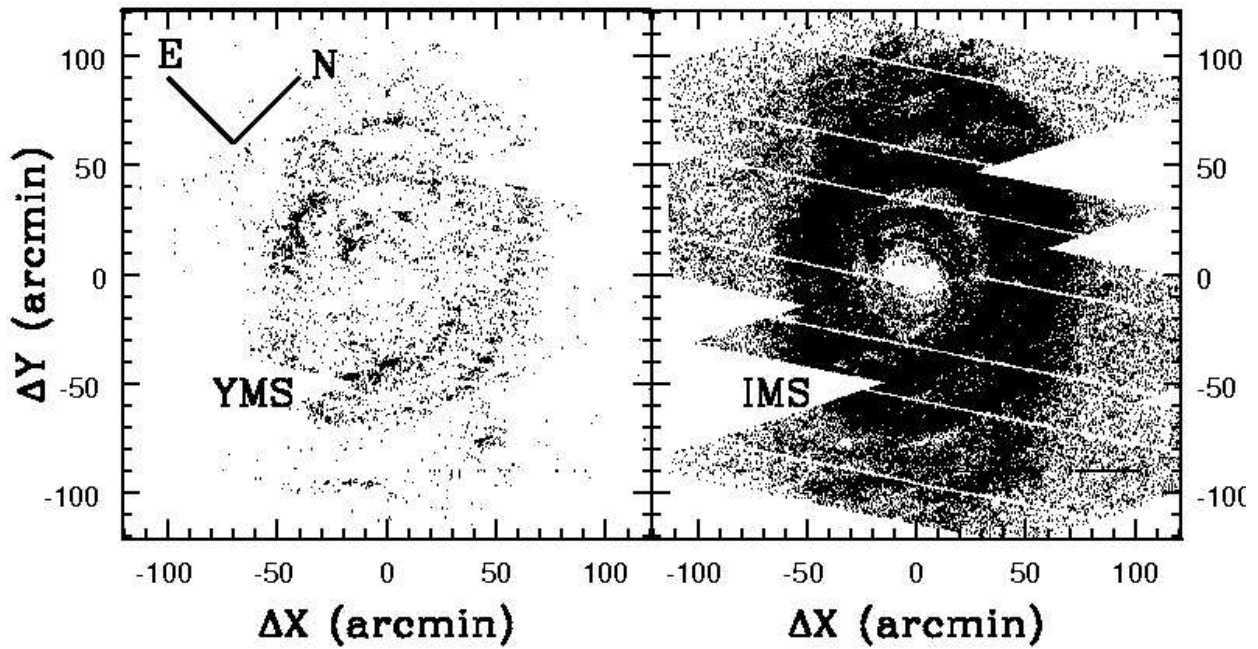}
\caption{The same as Figure 17, but showing the distribution of objects as 
they would appear if M31 were viewed face-on. The orientation of M31 has been 
rotated so that the semi-major axis is vertical. North and East are indicated.}
\end{figure}

\clearpage

\begin{figure}
\figurenum{19}
\epsscale{0.75}
\plotone{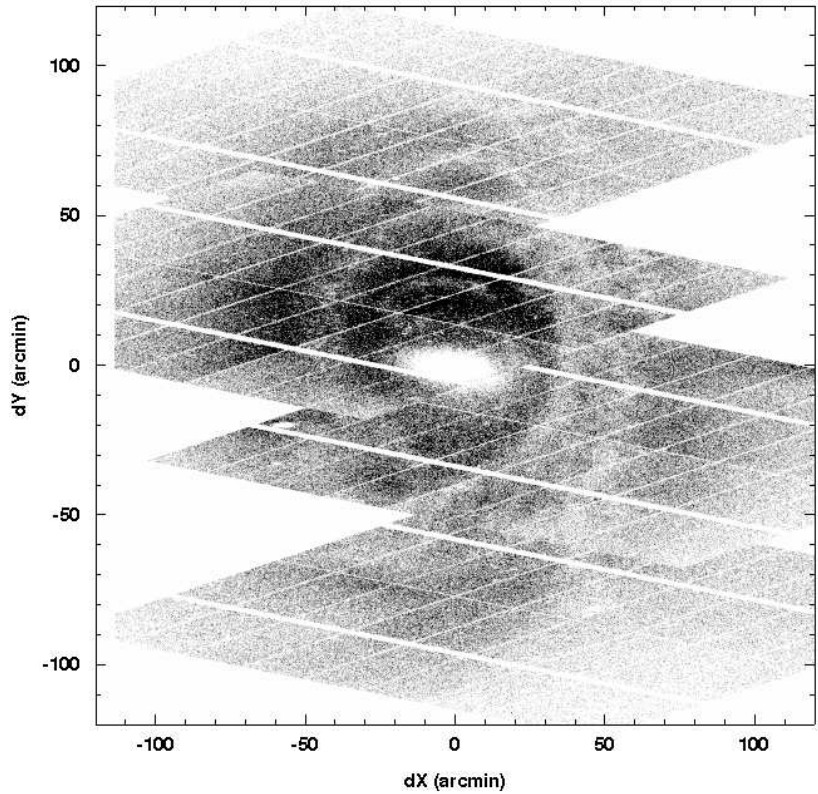}
\caption{The distribution of sources in the RHeB$+$AGB sample, as they would 
appear if M31 were viewed face-on. The major axis of M31 points 
upwards. The intensity at each location reflects number counts in $6 \times 
6$ arcsec arcsec bins. The source distribution is asymmetric, 
in the sense that the density of objects above the galaxy center is 
much higher than to the south of the galaxy center. M32 is located 
near (dX,dY) = (--60,--20), while the concentration of sources near (dX,dY) 
= (100,0) is probably associated with NGC 205.}
\end{figure}

\clearpage

\begin{figure}
\figurenum{20}
\epsscale{0.75}
\plotone{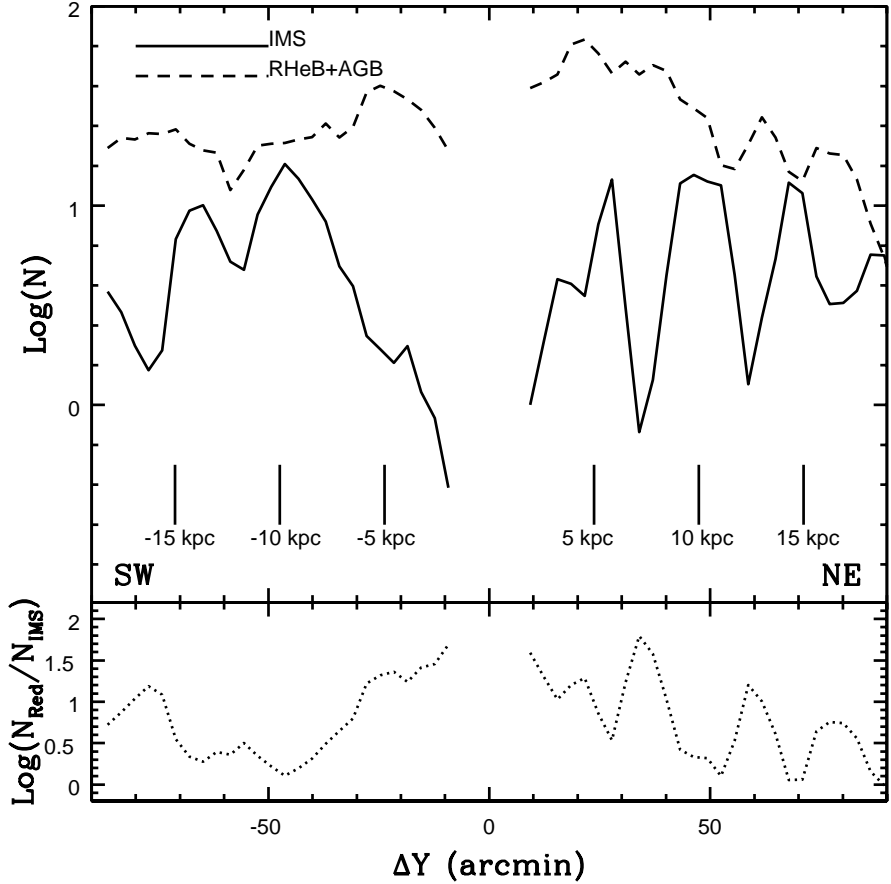}
\caption{(Top panel) The distribution of RHeB$+$AGB and IMS objects along the major 
axis of M31. $N$ is the number of stars arcmin$^{-2}$ in the de-projected dataset 
within $\pm 4$ arcmin of the major axis. Positive distances are along the NE 
axis, while negative distances are along the SW axis. The number density of 
evolved red stars is $\sim 0.4$ dex higher at R = 5 kpc than at R = --5 kpc; 
the distribution of these objects in M31 is thus lop-sided. 
The spatially periodic nature of recent star-forming regions is 
also clearly evident in the IMS profile along the NE axis. (Bottom 
panel) The ratio of stars in the RHeB$+$AGB and IMS samples. The peaks due to 
star-forming rings aside, there is a tendency for the ratio to decrease 
towards larger radii along the NE axis, indicating that the relative frequency 
of younger stars grows with increasing radius.}
\end{figure}

\end{document}